\documentclass[twocolumn]{aastex62}
\usepackage{amsmath}
\usepackage{graphicx}
\usepackage{soul}

\newcommand{\vmax}{\ensuremath{v_{\rm max}}}

\newcommand{\dproj}{\ensuremath{d_{\rm proj}}}
\newcommand{\dv}{\ensuremath{\Delta v_{\rm los}}}
\newcommand{\zwate}{{ZwCl 0008}}

\begin{document}
\title{Dynamical Properties of Merging Galaxy Clusters from Simulated Analogs}

\shorttitle{Dynamics of Merging Galaxy Clusters}

\author[0000-0002-0813-5888]{David Wittman}
\affiliation{Physics Department, University of California, Davis, CA 95616}
\email{dwittman@physics.ucdavis.edu}

\begin{abstract}
  Merging galaxy clusters may provide a unique window into the
  behavior of dark matter and the evolution of member galaxies.  To
  interpret these natural collider experiments we must account for how
  much time has passed since pericenter passage (TSP), the maximum relative
  speed of the merging subclusters, merger phase (outbound after first
  pericenter or returning for second pericenter), and other dynamical
  parameters that are not directly observable. These quantities are
  often inferred from staged simulations or analytical timing
  arguments that include neither substructure, large-scale structure,
  nor a cosmologically motivated range of impact parameters.  We
  include all these effects by extracting dynamical parameters from
  analog systems in a cosmological n-body simulation, and we present
  constraints for 11 observed systems.  The TSP and viewing angles we
  derive are consistent with those of staged hydrodynamical
  simulations, but we find lower maximum speeds. Compared to the
  analytical MCMAC method we find lower TSP, and viewing angles that
  put the separation vector closer to the plane of the sky; we
  attribute this to the MCMAC assumption of zero pericenter distance.
  We discuss potential extensions to the basic analog method as well
  as complementarities between methods.
\end{abstract}


\section{Introduction}\label{sec-intro}

The merger of two clusters of galaxies triggers a range of
astrophysical processes. Shocks in the intracluster medium launched
around the time of pericenter trigger synchrotron emission
\citep[detectable as radio
``relics'':][]{Ensslin1998,Feretti2012,Skillman13}, accelerate cosmic
rays \citep{Brunetti2014}, and may affect star formation and AGN
activity \citep{MillerOwen2003,Sobral2015}. X-ray morphology is also
greatly affected as gas associated with each cluster can be stripped
or displaced around the time of pericenter. Mergers also place upper
limits on momentum exchange between dark matter particles
\citep{Markevitch04,Randall2008,Kahlhoefer13,RobertsonBullet2017,Mismeasure2018,HarveyWobble2018}. Any
galaxy-dark matter displacement created at pericenter crossing can,
however, change sign later as the galaxies in each subcluster fall
back to, and through, the center of their host dark matter halo
\citep{Kim17}. Hence the interpretation of an observed state hinges
on knowing the merger phase (outbound toward first apocenter or
returning toward second pericenter) and more specifically the time
since pericenter (TSP).  Merger effects on star formation must be
interpreted in the same light. The claims that some mergers have
stimulated star formation
\citep{MillerOwen2003,MaEbelingMarshall2010,Sobral2015,Stroe2017},
some have quenched it \citep{Mansheim2017}, and some have had no
effect \citep{Chung2010} may not conflict given that each
system is seen at a different TSP. It is also possible that pericenter
speed has an effect with low speeds compressing, but high speeds
disrupting, star-forming gas.  A coherent picture can emerge only if
we have access to robust methods of inferring the dynamics of each
system from the one snapshot we observe.

Efforts to do this date back to the timing argument of
\citet{Kahn1959}, which uses equations of motion for two point masses
on radial trajectories in an expanding universe.  Given the observed
masses, separation, and relative speed, the timing argument can reveal
the time since pericenter and speed at the time of
pericenter. Observations, however, reveal only the projected
separation and line-of-sight component of the relative velocity
vector. This motivated \citet{Dawson2012} to update the timing
argument with a code,
MCMAC,\footnote{\url{https://github.com/MCTwo/MCMAC}} that
marginalizes over possible viewing angles, along with other
improvements such as using Navarro-Frenk-White \citep{NFW97} mass
profiles rather than point masses.  He showed that uncertainty in
viewing angle is a substantial source of uncertainty in dynamical
parameters such as time since pericenter (TSP) and the maximum speed
\vmax.

\citet[][hereafter WCN18]{Analogs2018} constrained the viewing angle
of observed clusters by ``observing'' analogs in the MultiDark
cosmological n-body simulation.  This work improved the viewing angle
constraints as follows. Because MCMAC assumes radial orbits, the
velocity vector is always parallel to the separation vector; thus any
nonzero observed line-of-sight velocity difference between
subclusters, $|\dv|>0$, rules out the possibility of the separation
vector being in the plane of the sky. This constraint is noteworthy
because in its absence spherical geometry would dictate that the sky
plane is the {\it most} likely orientation for a random vector.  WCN18
showed that in mergers drawn from cosmological n-body simulations the
velocity vector does have some component perpendicular to the
separation vector, so that the plane-of-sky configuration remains
quite likely for $|\dv|$ up to several hundred km/s.  Most observed
systems have $|\dv|$ in this range \citep{MCCsampleanalysis}.  Indeed,
WCN18 found that for most of the systems they considered, the
likelihood of the viewing angle (defined as
the angle between the line of sight and the separation vector) is a
monotonically rising function that peaks at $90^\circ$, whereas MCMAC
typically produced a rise followed by a sharp cutoff from
$\approx80-90^\circ$. The two methods did agree that observed systems
with large $|\dv|$ ($\gtrsim 1000$ km/s) have separation vectors
$\sim45^\circ$ from the line of sight; in other words, a substantial fraction
of a subcluster's plunging motion must be along the line of sight to
make $|\dv|$ this large.

In this paper we extend the analog method to infer the dynamical quantities
of most interest, TSP and \vmax. We also show that for some observed
systems analogs can reveal the phase of the orbit, i.e. whether the
subclusters are outbound after first pericenter or returning after
first turnaround. 

The remainder of the paper is organized as follows.  In
\S\ref{sec-method} we outline our method, and in \S\ref{sec-results}
we show results for nine merging systems with a range of properties.
In \S\ref{sec-mcmac} we compare our procedure with MCMAC by finding
analogs of the two systems analyzed by \citet{Dawson2012}, and in
\S\ref{sec-discussion} we summarize and discuss the implications.  To
maintain consistency with the simulation described below, we adopt the
flat \citet{PlanckCosmology2014} cosmology, in which $H_0=67.8$
km/s/Mpc and $\Omega_m=0.307$.

\section{Method}\label{sec-method}

As in WCN18, we used the publicly available Big Multidark Planck
(BigMDPL) Simulation \citep{BigMDPL2016} hosted on the
\textit{cosmosim.org} website.  BigMDPL has a box size of (2.5
Gpc/h)$^3$ and a particle mass of $2.359 \times 10^{10} M_{\odot}$,
yielding at least 2500 particles for all halos in the mass range we
consider. The database includes a halo catalog created by the Rockstar
algorithm \citep{Rockstar2013}, from which we extract all halos with
virial masses $>6 \times 10^{13}\,M_{\odot}$. We then find pairs of
halos separated by $\le 5$ Mpc, excluding pairs for which either
member is within 5 Mpc of a third halo. This is to mimic the selection
of observed binary clusters; the method could be extended to select
analogs of more complicated mergers but this is beyond the scope of
this paper. Having done this separately for all snapshots, we then
match pairs across snapshots to obtain a history for each pair and, as
in WCN18, discarded pairs that were never separated by $<300$ kpc in
any snapshot as well as pairs with multiple pericenters closer than
this.  This is because the real clusters for which we seek analogs
have stripped X-ray morphologies that strongly suggest a recent first
pericenter with a small pericenter distance.

\begin{figure}
\centerline{\includegraphics[width=0.96\columnwidth]{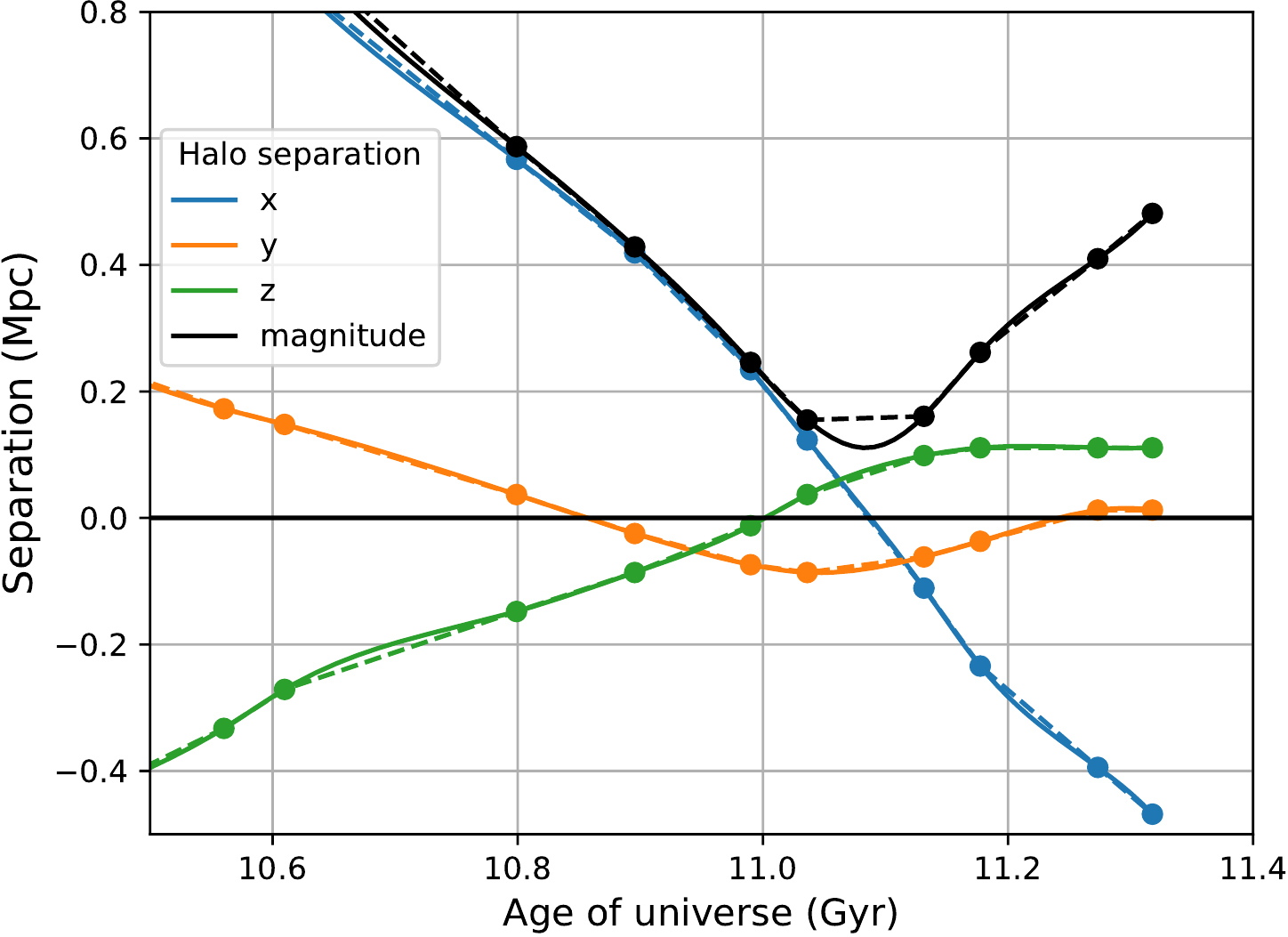}}
\centerline{\includegraphics[width=\columnwidth]{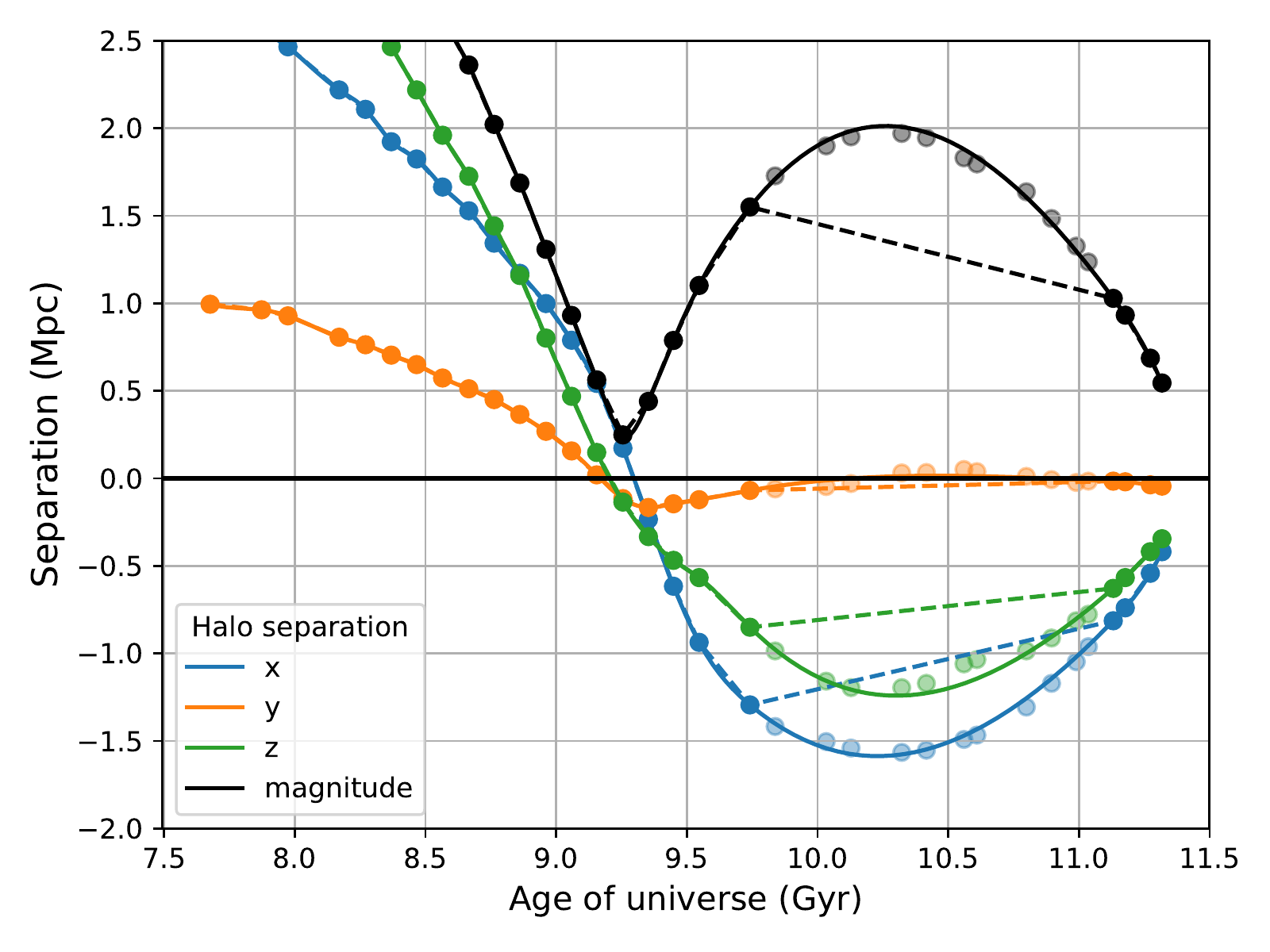}}
\caption{Interpolation performance: markers indicate halo separations
  recorded at the snapshots, with dashed (solid) curves indicating
  linear (cubic spline) interpolation. {\it Top:} linear interpolation
  of the separation magnitude (dashed black line) overestimates
  pericenter distance, so we use the magnitude of the interpolated
  values of the more smoothly varying cartesian components (black
  curve). This works well regardless of the type of interpolation used
  for the components. {\it Bottom:} the interpolation type
    matters when there are gaps in the halo catalog time series.
    Here, translucent points withheld to simulate such a gap
    demonstrate the accuracy of the cubic spline interpolation.}
\label{fig-interp}\end{figure}

We selected one halo from each pair to serve as a reference, and
recorded the separation and velocity vectors at each snapshot.  This
improves on the method of WCN18, who recorded only the magnitude of
the separation at each snapshot, as follows.  Unlike WCN18 we are
concerned with the time evolution and wish to interpolate between
snapshots.  The cartesian components of the halo separation vector
vary smoothly through pericenter while the magnitude does not, as
shown in the top panel of Figure~\ref{fig-interp}. Hence we
interpolate the components onto a finer time grid and use these
interpolated components to reconstruct the magnitude of the
separation. The top panel shows that linear interpolation of the
magnitude overestimates the pericenter distance, and better
performance is provided by either linear or cubic spline interpolation
of the components.  The bottom panel shows that when there are gaps,
cubic spline is a better way to interpolate the components.  We also
tried cubic spline interpolation of the magnitude, but this caused
unphysical ringing.

In a few trajectories, the halo separation recorded in the BigMDPL
database appears not to be a smooth function of snapshot number. This
may be due to accretion of smaller halos or perhaps to artifacts of
assigning particles to halos. In these cases the interpolation causes
a small amount of ringing evident as ripples in a few of the trajectories
presented below. This does not substantially affect the inference of
TSP or pericenter distance, because the ripples are small and because
they occur in a small fraction of trajectories. We retain the cubic
spline interpolation in spite of these occasional ripples because it
works across gaps (Figure~\ref{fig-interp}, bottom panel). This is not
strictly necessary for this paper, which deals only with the times of
observation and of pericenter, but it potentially enables studies of
apocenter distances and periods.

In about 5\% of the trajectories, the halo catalogs had swapped the
halo identification numbers at some point near pericenter; this was
evident by the sudden reversal of sign of all relative position and
velocity components.  This leaves a false impression that each
component passed through zero between snapshots, but it does not
change the separation magnitude at the snapshots. Hence, linear
interpolation of the separation magnitude is more robust in these
cases.  We were able to detect and undo most of the halo swaps in an
automated way by triggering on the change in the velocity vector from
one snapshot to the next; halo swaps make this quantity unphysically
large. Nevertheless, there were still some cases (often involving
multiple halo swaps) where an attempted unswapping did not
overwhelmingly improve this metric, and in these cases ($\approx 2\%$
of the total) we fell back to the less precise but more robust
interpolation of the separation magnitude.

% time res; relates to gaps 
BigMDPL snapshots are typically separated by redshift increments of
about $0.01$, or time increments of about 100 Myr, so our
interpolation typically yields TSP accurate to $\approx 10$ Myr. This
is a small fraction of the typical cluster TSP of $\approx 400$
Myr. At some redshifts, however, BigMDPL leaves a larger
($\approx 0.04$) redshift gaps between snapshots. In the few cases
where an observed cluster fell at one of these redshifts, we
``observed'' the analogs at a slightly higher redshift so the
preceding trajectory would be well sampled.

\begin{figure}
\centerline{\includegraphics[width=\columnwidth]{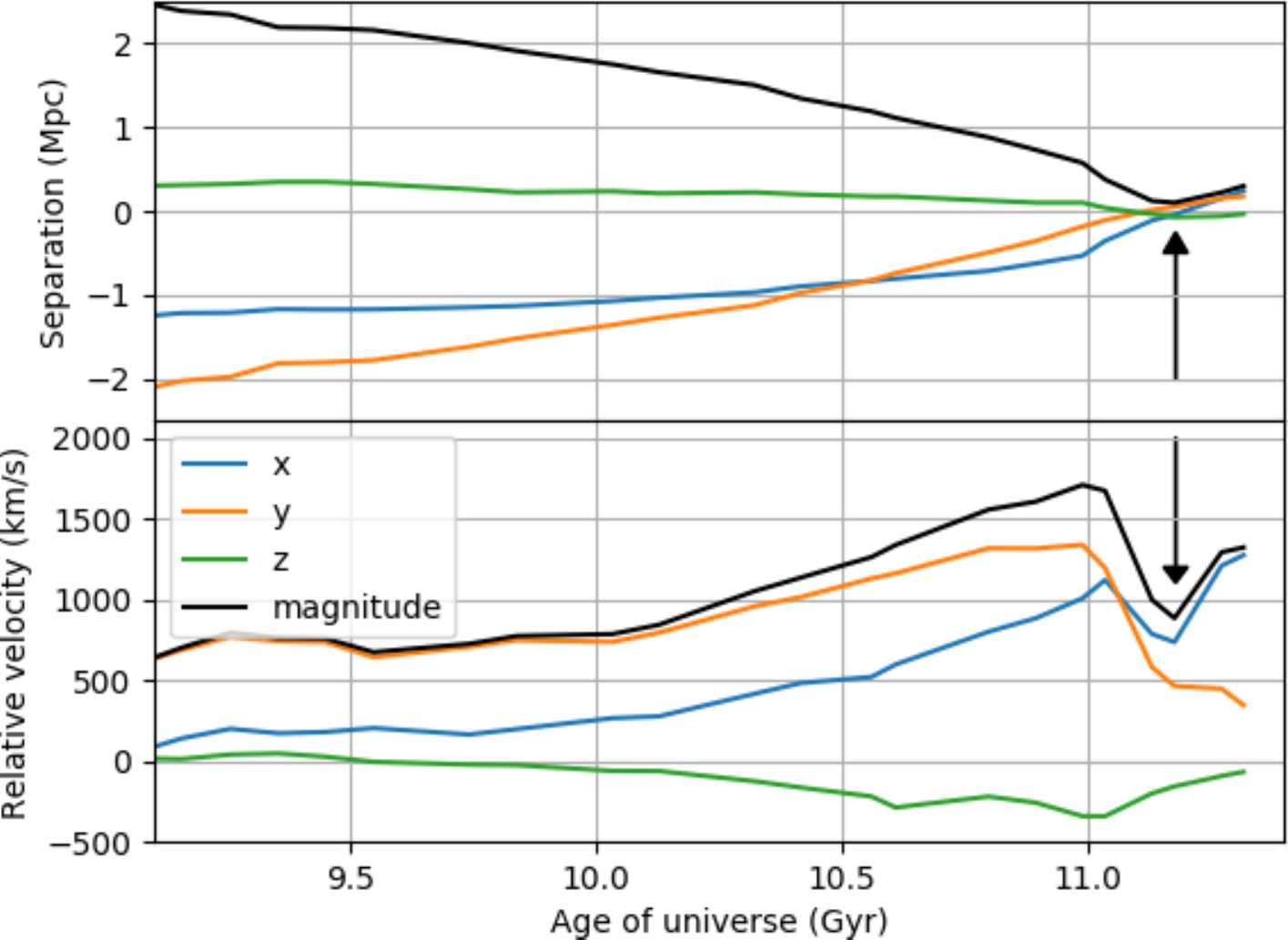}}
\caption{An extreme example of the drop in the halo relative velocity
  (lower panel) before the separation (upper panel) reaches a minimum
  (at the time indicated by the arrows). The components of each vector
  are shown in color and the magnitude in black. This drop in speed is
  likely an artifact due to difficulty in assigning particles to
  halos. Therefore, we record the maximum speed rather than the speed
  at the time of pericenter. }
\label{fig-vdrop}\end{figure}

% vmax
This procedure gives good resolution for TSP and pericenter distance, but the
velocity history is not so easily interpolated. We found that the
halo catalogs often recorded a drop in relative velocity around the
time of pericenter; Figure~\ref{fig-vdrop} shows an extreme
example. This is presumably an artifact of identifying which particles
belong to which halo. (Such artifacts may exist in the halo positions
as well, but the separation at this time is so small that the absolute
bias in separation cannot be large; in contrast, the velocity should
be maximum here so the bias is notable.) Because the cataloged speed
at the time of pericenter could be a substantial underestimate, we
tabulate \vmax, the maximum cataloged speed, which typically occurs
before pericenter.  This is still a slight underestimate of the
pericenter speed: a linear extrapolation\footnote{The speed is rising
  approximately linearly with time because these are extended halos,
  not point masses.}  of the rising speed to the time of pericenter
typically yields 100--200 km/s additional speed.

The probability of an analog matching the observed masses, \dproj, and
\dv\ at a given polar viewing angle $\theta$ is calculated as in WCN18.
Note that this calculation is unaffected by any halo swap, as it
depends only on the ``observed'' snapshot rather than the time
evolution.  We rewrote the WCN18 code for this and most other steps to
be more efficient, and we verified that both codes deliver the same
result on common outputs such as the probability as a function of
viewing angle. Here, however, we integrate over the viewing angle to
assign an overall probability to each analog. This probability is then
used to weight the TSP of that analog when inferring TSP, and
similarly with the other dynamical parameters.

{\it Parameter recovery tests.}  We present a series of tests
  using BigMDPL systems as ``observed'' systems where the true values
  of the parameters are known. We use systems from snapshot 74
  (corresponding to redshift 0.1058) and assign the observational
  uncertainties listed for the observed cluster ZwCl 0008.8+5215
  below. In each case we remove the ``observed'' system from the
  BigMDPL catalog, run the analog-finding code, and compare the
  results to the true values for the system.

For each system we infer TSP, \vmax, viewing angle $\theta$ (with the
convention that $\theta=90^\circ$ corresponds to a separation vector
in the plane of the sky), and the angle $\varphi$ between the
separation vector and the velocity vector of the analog.  This serves
as a proxy for outbound vs. returning phases of the merger:
$\varphi=0$ indicates radially outbound, $\varphi=180^\circ$ indicates
radially inbound, and nonradial trajectories will shift from
$\varphi<90^\circ$ to $\varphi>90^\circ$ at apocenter.  We do not
attempt to infer the pericenter distance because we found a similar
broad range of pericenter distances regardless of the specific values
of \dproj\ and \dv\ used to winnow the analogs.  It is possible that
other observables such as X-ray morphology could constrain the
pericenter distance if the analog method were to be extended to
hydrodynamic simulations.

We begin with System A, whose distinguishing feature is a low 3-D
velocity at the time of observation ($v_{3D}(t_{\rm obs}) = 534$ km/s,
in the bottom 2\% of systems in snapshot 74) which will make \dv\
change little with viewing angle.  This is coupled to a large TSP (802
Myr, in the top 10\% of systems identified in snapshot 74).  With a
separation vector of $(0.574,0.210,-0.818)$ Mpc and a velocity vector
of $(376,164,-342)$ km/s, placing the line of sight along the $y$ axis
yields $\dv=164$ km/s and $\dproj=0.999$ Mpc, typical of observed
systems considered in this paper.

\begin{figure*}
\centerline{\includegraphics[width=\textwidth]{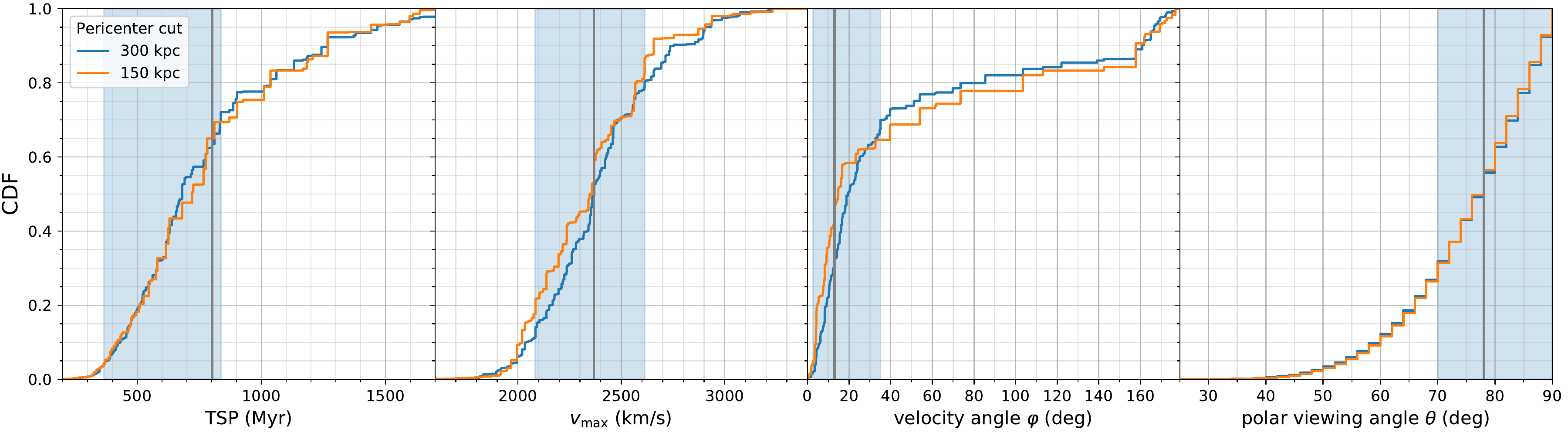}}
\caption{CDFs of dynamical parameters inferred from analogs of System
  A, with blue curves for all analogs passing the initial 300 kpc cut
  on pericenter distance, and orange for the subset passing a stricter
  cut of 150 kpc.  The similarity of the curves demonstrates that
  dynamical parameters can be inferred without prior knowledge of the
  pericenter distance.  The true values (gray lines) are in all cases
  within the 68\% HPDCI (shaded regions) and also close to where the
  CDF crosses 0.5, which marks the median (likelihood-weighted) analog
  value. In most panels, each step in the CDF represents the
  contribution of one analog. In the $\theta$ panel, the 2$^\circ$ steps 
  reflect the sampling of lines of sight.}
\label{fig-test062}\end{figure*}

Figure~\ref{fig-test062} shows the cumulative distribution functions
(CDFs) for the inferred parameters using the default set of analogs
(blue curves) and a restricted set with small ($<150$ kpc) pericenter
distance (orange curves). The two sets of curves are quite similar,
indicating that these parameters can be inferred without knowledge of
the pericenter distance.  We further tested the effects of this cut on
other simulated systems, with consistently negligible effects.  We
therefore focus on the blue set of curves in the following comments on
Figure~\ref{fig-test062}, and retain the 300 kpc cut for the remainder
of this section.  The following section will show that results for
real observed systems are also insensitive to this cut.

The shaded regions in Figure~\ref{fig-test062} mark the 68\% highest
probability density confidence intervals (HPDCI). In other words, each
blue curve exits this region 0.68 higher than it entered, and the
region shown is the most compact contiguous region meeting this
criterion.  In each panel, the true value (gray line) is within the
68\% HPDCI, and also near the median analog value (where the curve
crosses 0.5). This shows that the inference procedure works well, at
least for this merging system and this line of sight.

We now examine how the constraints may vary with the viewing
geometry. Table~\ref{tab-los} lists the basis vectors for five lines
of sight (LOS) for System A. LOS 2 is the one already examined above,
while the others explore a range of \dproj\ and \dv. Note that the
relationship between \dproj\ and \dv\ is not strictly inverse; this is
because the relative velocity vector is not parallel to the separation
vector.

\begin{table}
\begin{tabular}{ccccr}
Observer & LOS & $\theta$ (deg)&  \dproj\ (Mpc)& \dv\ (km/s)\\ \hline
1&$(1,0,0)$ &   56  &   0.845 &  56\\
2&$(0,1,0)$ &  78  &   0.999 &164\\
3&$(0,0,1)$ &   37   &  0.611& 342\\
4&$(1,0,1)$ &   80    & 1.006& 24\\
5&$(1,0,-1)$ & 15    & 0.272& 508\\
\end{tabular}
\caption{Lines of sight of mock observers of System A.}
\label{tab-los}
\end{table}

\begin{figure*}
\centerline{\includegraphics[width=\textwidth]{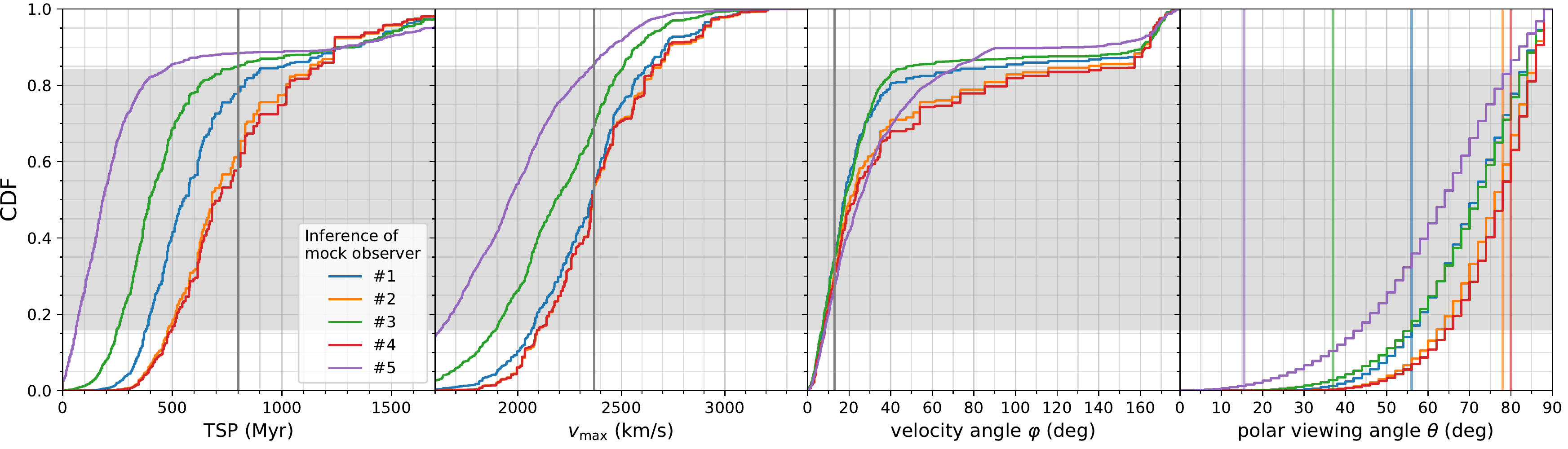}}
\caption{CDFs for dynamical parameters from analogs of System A, for
  the five different lines of sight and mock observables listed in
  Table~\ref{tab-los}. The true values (vertical gray lines) are
  generally within the central 68\% CI (shaded regions), but Observers
  3 and 5 underestimate TSP and overestimate $\theta$ due to the
  combination of an atypical system with atypical viewing angles as
  explained in the text.}
\label{fig-Alice5}\end{figure*}

Figure~\ref{fig-Alice5} shows the constraints for the five LOS coded
by color. For clarity, the figure shows the central 68\% confidence
interval (CI) rather than multiple individually tailored HPDCI.  The
first two panels show that the constraints on TSP and \vmax\ become
less accurate as the separation vector approaches the LOS
($\theta=15^\circ$ for Observer 5 and $37^\circ$ for Observer
3). Nevertheless, the constraints are still workable for Observer 3:
the true \vmax\ is well within the 68\% CI and the true TSP is well
within the 90\% CI.  The third panel shows that the $\varphi$
inference is stable across the various LOS.  The fourth panel is the
only panel in which the true value depends on the LOS, and it shows
that the $\theta$ CDF does not change as much as we would like in
response to real changes in the viewing angle.  This is because System
A has a low $v_{3D}(t_{\rm obs})$, which is only 3.4 times the
observational uncertainty. Hence even the extreme $\theta=0$ and
$\theta=90^\circ$ models offer predictions for \dv\ that conflict only
mildly, which limits the ability to distinguish models. As a result,
the $\theta$ CDF hews closely to that predicted for random observers
in the absence of data. This causes Observers 3 and 5 to place the
separation vector too close to the plane of the sky, thus
underestimating the current 3-D separation given their \dproj. This in
turn underestimates TSP, the time required to reach that 3-D
separation.

\begin{figure*}
\centerline{\includegraphics[width=\textwidth]{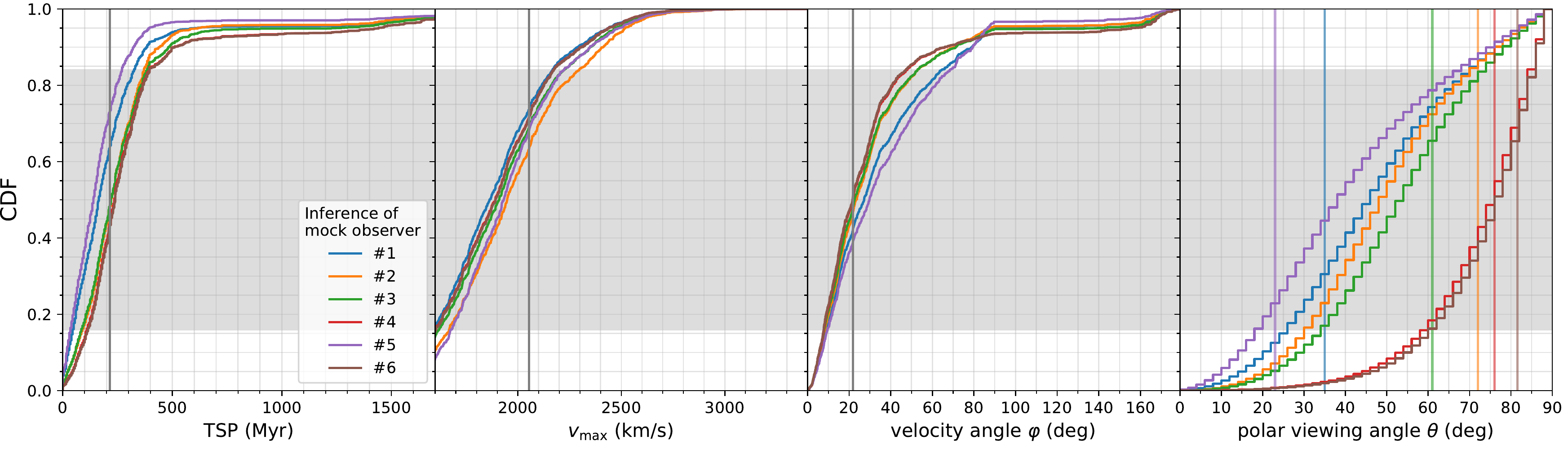}}
\caption{CDFs for dynamical parameters from analogs of System B, for
  the six lines of sight listed in Table~\ref{tab-losB}. The true values
  (vertical gray lines) are all well within the central 68\% CI
  (shaded regions).}
\label{fig-dawn}\end{figure*}

We now contrast this with System B, which has
$v_{3D}(t_{\rm obs})= 1746$ km/s, slightly above the snapshot 74
average of 1437 km/s, in part because it is observed only 215 Myr
after pericenter. Table~\ref{tab-losB} lists the data for the same
five mock observers plus a sixth observer tailored to produce very low
$\dv$; Figure~\ref{fig-dawn} shows the inferred CDFs. With System B
the true values are more consistently well within the central 68\% CI.
Furthermore, inferences on TSP, \vmax, and $\varphi$ are much more
stable across the various LOS, while the inference on $\theta$ is much
more responsive to the LOS.  Still, $\theta$ tends to be overestimated
when its true value is low, and underestimated when its true value is
high. This is a feature of Bayesian inference: estimates are correct
on average for an ensemble of observers, but not for selected
observers atypical of the prior distribution.

\begin{table}
\begin{tabular}{ccccc}
Observer & LOS & $\theta$ (deg)&  \dproj\ (Mpc)& \dv\ (km/s)\\ \hline
1&$(1,0,0)$ &  35  &   0.238 &1052\\
2&$(0,1,0)$ &   72  &   0.394 & 1086\\
3&$(0,0,1)$ &   61   &  0.364& 874\\
4&$(1,0,1)$ &   76    & 0.403& 126\\
5&$(1,0,-1)$ & 23    & 0.163& 1362\\
6&$(0.82,0,1)$ & 82    & 0.411& 9\\
\end{tabular}
\caption{Lines of sight of mock observers of System B.}
\label{tab-losB}
\end{table}

In practice, biases as large as that of Observer 5 on System A will be
exceedingly rare: only 3.4\% of random observers have
$\theta\le15^\circ$, and such observers will not identify the system
as bimodal to begin with due to their small \dproj\ {\it and} small
\dv.  The bias also fades for systems with more typical
$v_{3D}(t_{\rm obs})$: compare the moderate TSP bias of System A's
Observer 3 ($\theta=37^\circ$) with the negligible TSP bias of System
B's Observer 5 ($\theta=23^\circ$).  A real-life example similar to
the low-$\theta$ observers of System B is RXC J1314.4-2515, which has
$\dv$ above 1000 km/s and for which WCN found (and we confirm)
$\theta\approx45^\circ$. This supports the contention that the
$\theta$ inference responds to the data where possible while
staying closer to a random-observer distribution where it must.

The larger pattern here is, for lack of a better term, regression to
the mean. The mean $v_{3D}(t_{\rm obs})$ of systems in snapshot 74 is
1437 km/s, and System A is in the bottom 2\% of systems in this
regard. Hence an inference using all systems will necessarily pull the
model $v_{3D}(t_{\rm obs})$ upward {\it in this case}, leading to an
upward bias in $\theta$ (to account for the small observed \dv) which
causes a downward bias in the 3-D separation and hence in TSP.
However, these effects will reverse for systems
with unusually high $v_{3D}(t_{\rm obs})$ so this is
not a bias {\it per se} but simply the application of cosmologically
motivated priors.

\section{Results}\label{sec-results}

This section begins by contrasting two clusters in detail to highlight
the interpretion of the figures and to point out some trends. Then, we
show results for each cluster in the \citet{MCCsampleanalysis} gold
sample but with less commentary on each cluster. 

\subsection{An Illustrative Pair of Systems: ZwCl 0008.8+5215 and MACS J0025.4--1222}\label{sec-illpair}

We first look at two systems in detail to illustrate what can be
learned from the analogs. We choose ZwCl 0008.8+5215 (hereafter
\zwate) and MACS J0025.4--1222 (hereafter MACS J0025) because they are
similar in some respects ($\dv\approx 100$ km/s and similar masses)
while differing in \dproj\ (1057 kpc for the former, but only 541 kpc
for the latter). They also differ in redshift, with $z=0.10$ and 0.59
respectively.  The values for \zwate\ and MACS J0025 in
Table~\ref{tab-clusterinputs} are adopted from
\citet{MCCsampleanalysis} and \cite{bradac2008} respectively, with
their angular separation values converted to physical using our
adopted cosmology.  Table~\ref{tab-clusterinputs} lists the values
used for all clusters in this paper.

\begin{table*}
\centering 
\begin{tabular}{lccccc}
Cluster & z & $M_1(10^{14} M_\odot)$ & $M_2(10^{14} M_\odot)$ & $d_{\rm proj} (Mpc)$ & $\Delta v_r (km/s)$ \\
\hline 
\multicolumn{6}{c}{Merging Cluster Collaboration Gold Sample\tablenotemark{a}}\\
\hline 
Abell 1240\tablenotemark{b} & 0.19 & 4.19 (0.99)  &4.58 (1.40) & 1.020 (0.201) & 395 (230)\\
Abell 3411\tablenotemark{b} & 0.16 & 9.0 (5.0) & 7.0 (5.0) & 1.286 (0.173) & 141 (195)\\
CIZA J2242.8+5301\tablenotemark{c}& 0.19 & 11.0 (3.7) &9.8 (3.8) & 1.203 (0.194) & 385 (299)\\
MACS J1149.5+2223\tablenotemark{d}& 0.54 & 8.35 (1.3) &5.45 (3.4) & 0.972 (0.394) & 228 (281)\\
MACS J1752.0+4440\tablenotemark{b}& 0.36 & 13.22 (3.14)& 12.04 (2.59) & 1.177 (0.314) & 136 (186)\\
RXC J1314.4-2515\tablenotemark{b} & 0.25 & 6.07 (1.8) &7.17 (2.8) & 0.571 (0.239) & 1498 (293) \\
ZwCl 0008.8+5215\tablenotemark{e} & 0.10 & 5.7 (2.8)& 1.2 (1.4) & 1.057 (0.119) & 110 (155) \\
ZwCl 1856.8+6616\tablenotemark{b} & 0.30 & 9.66 (4.06) &7.6 (4.05) & 0.955 (0.278) & 227 (317) \\
\hline 
\multicolumn{6}{c}{Other Clusters}\\
\hline 
MACS J0025.4--1222\tablenotemark{f} & 0.59 & 2.5 (1.7) &2.6 (1.4) & 0.541 (0.102) & 100 (80)\\
DLSCL J0916.2+2951\tablenotemark{g} & 0.53 & 3.1 (1.2) &1.7 (2.0) & 1.030 (0.144) & 670 (330)\\
1E 0657-558\tablenotemark{h} & 0.30 & 15.0 (1.5) & 1.5 (0.15) & 0.742 (0.021) & 616 (62)\\
\end{tabular}
\caption{Observed values used for each cluster, with uncertainties in parentheses.}
\label{tab-clusterinputs}
\tablenotetext{a}{\citet{MCCsampledata,MCCsampleanalysis}}
\tablenotetext{b}{We modify the dynamical mass adopted by
  \citet{Analogs2018} as explained in \S\ref{sec-gold}.}
\tablenotetext{c}{\citet{JeeCIZA2014,DawsonCIZA2014}}
\tablenotetext{d}{\citet{GolovichMACS1149}; we modify their dynamical
  mass as explained in \S\ref{sec-gold}.}
\tablenotetext{e}{\citet{GolovichZwCl0008}}
\tablenotetext{f}{\cite{bradac2008}}
\tablenotetext{g}{\cite{Dawson2012,Dawson11}}
\tablenotetext{h}{\cite{Dawson2012,Barrena2002Bulletredshifts,BradacBulletLensing2006}}
\end{table*}

% convert 'ZwCl0008.pdf[1]' -flatten -trim +repage ZwCl0008-trajectories.pdf
\begin{figure}
\centerline{\includegraphics[width=\columnwidth]{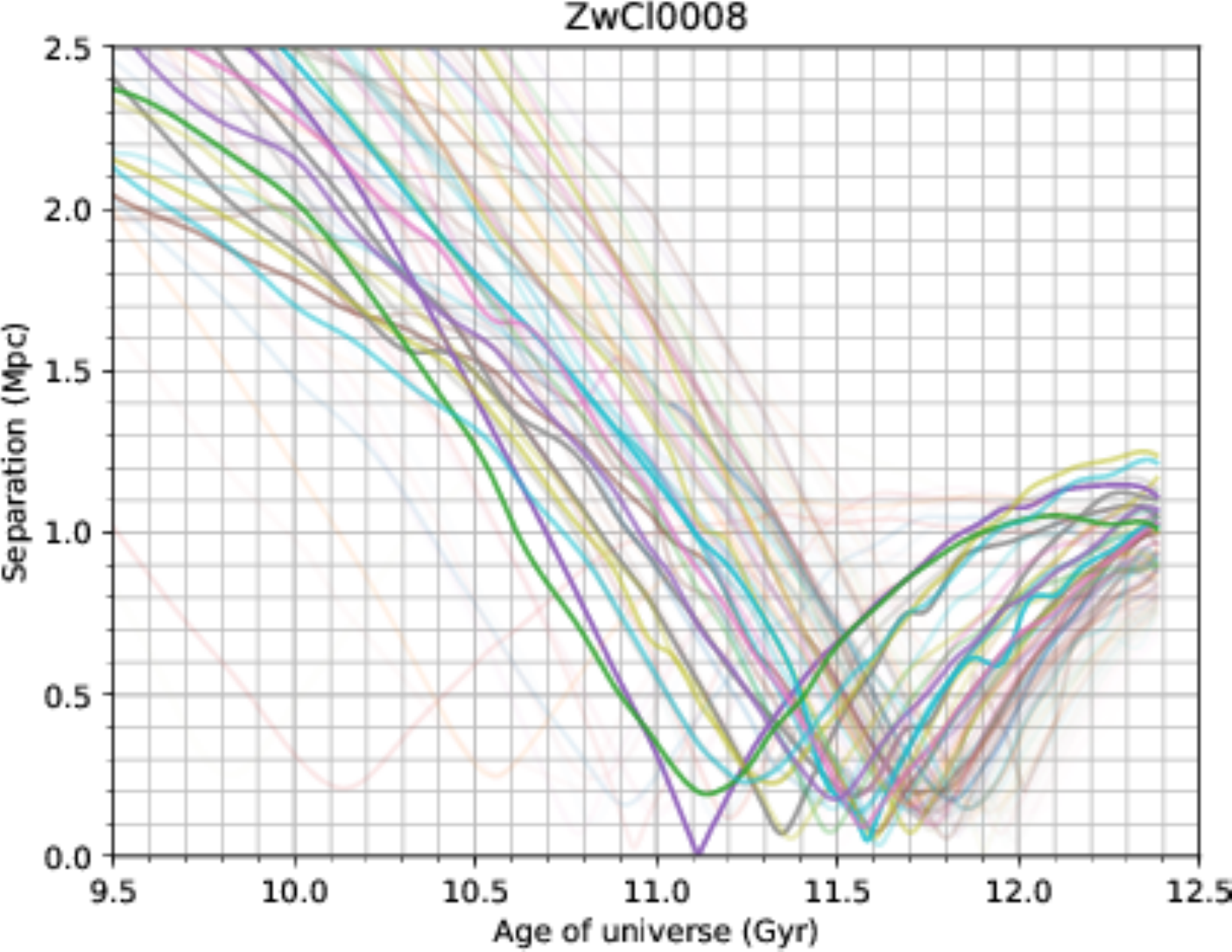}}
\caption{Physical separation versus time for analogs of ZwCl
  0008.8+5215. The opacity is proportional to the likelihood of
  the observed data given the analog, and colors are cycled to better
  distinguish individual analogs.}
\label{fig-ZwCltrajectories}\end{figure}

% convert 'MACSJ0025.pdf[1]' -flatten -trim +repage MACSJ0025-traj-crop.pdf
\begin{figure}
\centerline{\includegraphics[width=\columnwidth]{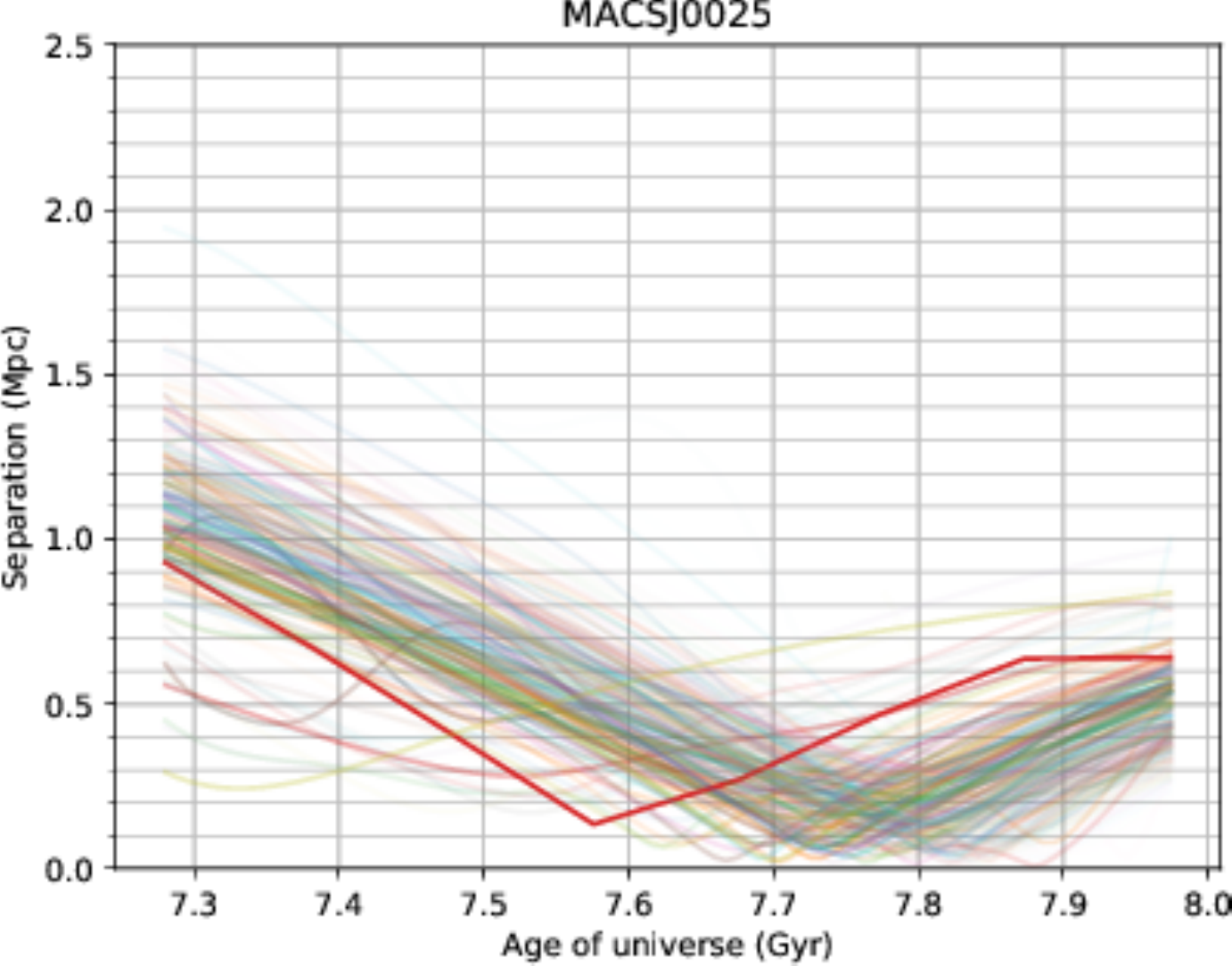}}
\caption{Physical separation versus time for analogs of MACS
  J0025.4--1222. The opacity is proportional to the likelihood of the
  observed data given the analog, and colors are cycled to better
  distinguish individual analogs.}
\label{fig-MACStrajectories}\end{figure}

Figures~\ref{fig-ZwCltrajectories} and \ref{fig-MACStrajectories} show
the 3-D separation versus time for analogs of \zwate\ and MACS J0025
respectively, with opacity encoding the likelihood of that analog
matching the observables in the final snapshot shown (which is the
time of observation). The small minority of trajectories that exhibit
ringing from the interpolation still have accurate pericenter
distances and times as explained in \S\ref{sec-method}. Note that the
two figures share a common separation scale but not a time scale. This
is because \zwate\ analogs are observed at a later snapshot, hence can
be traced further back in time. The MACS J0025 trajectories are shown
back to the start of dense time sampling in the BigMDPL database, 7.21
Gyr since the Big Bang, which is only 0.7 Gyr before the time of
observation.  \zwate\ analogs, in contrast, can be traced many Gyr
back (for clarity the figure truncates at 3 Gyr before observation).

The next most salient difference between
Figures~\ref{fig-ZwCltrajectories} and \ref{fig-MACStrajectories} is
that \zwate\ analogs seem to be falling from greater distances and
rebounding to greater distances by the time of observation.  The
larger infall distance is mostly an artifact of tracing \zwate\
analogs further back in time, while the larger separation at the time
of observation is driven by the larger observed projected separation.
There {\it are} some trajectories in Figure~\ref{fig-MACStrajectories}
with larger separations at the time of observation, but their
transparency indicates they are poor matches to the observations. To
some extent this is because their 3-D separations substantially exceed
the observed \dproj\ of 541 kpc in MACS J0025. This alone is not
enough to rule out such analogs, because projection effects can always
reduce a large 3-D separation to a small \dproj. However, by placing
the separation vector more parallel to the line of sight, most such
models will predict a large \dv\ that yields conflicts with the low value
observed for MACS J0025.

\begin{figure}
\centerline{\includegraphics[width=\columnwidth]{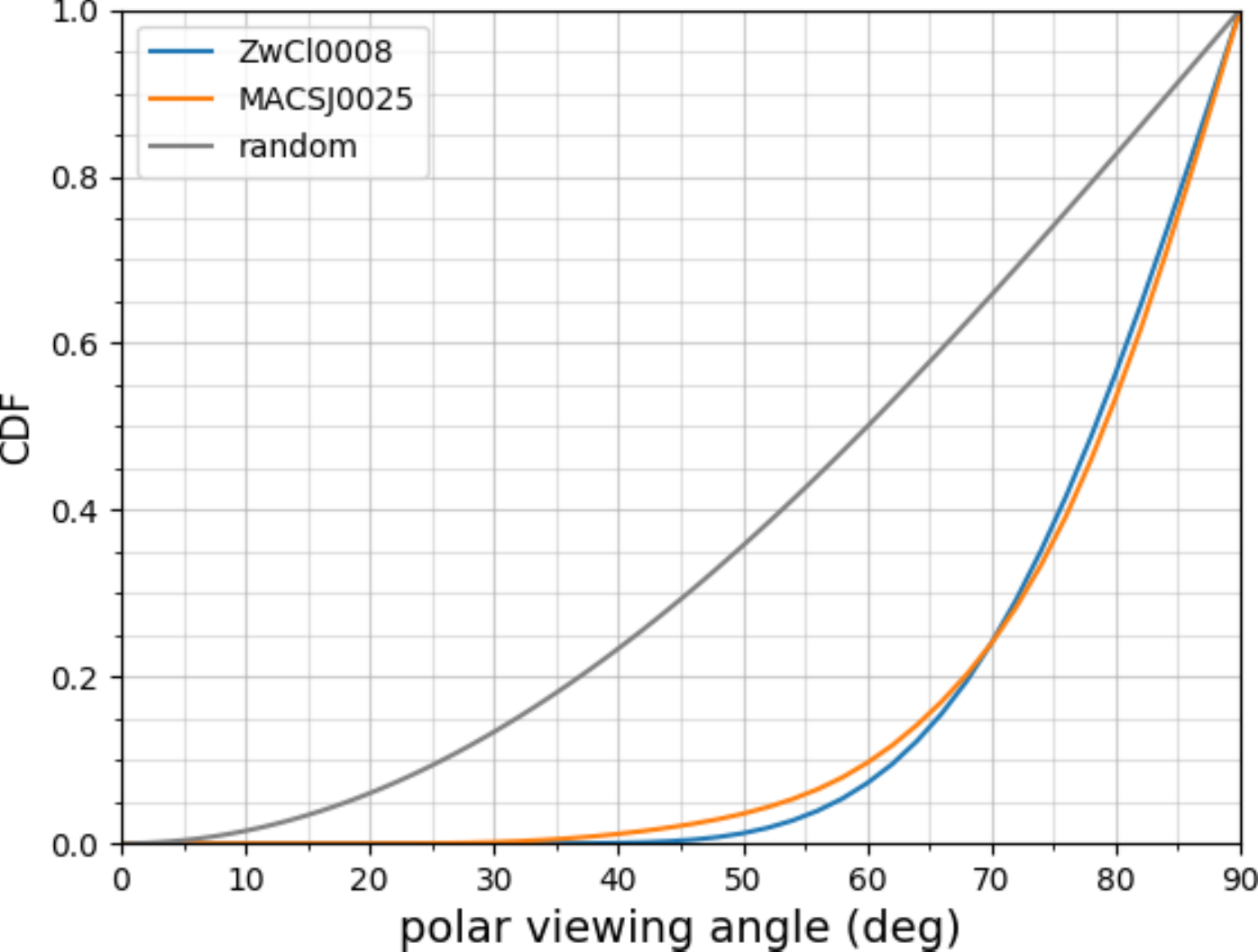}}
\caption{Viewing angle constraints for ZwCl 0008.8+5215 and MACS
  J0025.4--1222. Here, the convention is that a viewing angle of
  90$^\circ$ corresponds to a separation vector in the plane of the
  sky. The gray curve shows the expected CDF for random viewing
  angles.}
\label{fig-theta}\end{figure}

This description of the inner workings is supported by the viewing
angle constraints in Figure~\ref{fig-theta}.  The constraints for the
two systems are similar: $\theta>63^\circ\,(61^\circ)$ for \zwate\
(MACS J0025) at 90\% confidence, or $73^\circ\,(73^\circ)$ at 68\%
confidence. In other words, the viewing angles necessary to reduce
\dproj\ to, say, half the 3-D separation are ruled out, and this can
be attributed largely to the low observed \dv.  Readers may note that
all viewing angles are possible, even with low observed \dv, if the
3-D relative velocity is low, i.e. the system is observed near
turnaround. Nevertheless, Figure~\ref{fig-theta} implies that
line-of-sight configurations in BigMDPL fail to satisfy all the
constraints simultaneously, at least for these two observed systems.

%convert ~/Figure_1.png -flatten -trim +repage viewangle.png

There is more to learn from Figures~\ref{fig-ZwCltrajectories} and
\ref{fig-MACStrajectories}. In both figures, note how analogs with
larger current separation tend to have larger TSP (which can be read
off by tracing a trajectory leftward from the right edge and locating
the minimum).  For \zwate, analogs with large current separations are
favored due to the large \dproj, and their pericenters occurred
$\sim800$ Myr ago, albeit with a wide range. Analogs to MACS J0025, in
contrast, are scattered across a much smaller and more recent range of
TSP.

Yet another dynamical feature implicit in
Figures~\ref{fig-ZwCltrajectories} and \ref{fig-MACStrajectories} is
the fraction of analogs that are returning after first apocenter (as
opposed to still outbound toward first apocenter).  For \zwate, a
nonnegligible minority of trajectories fit this description, but for
MACS J0025 none do. (The opaque red trajectory is close to apocenter,
but in the last snapshot shown it has a velocity vector consistent
with outbound, and a separation that increased 5 kpc from the prior
snapshot.)  Analogs in the returning phase have substantially more TSP
than those in the outbound phase, so when they are consistent with
observations they can substantially increase the average TSP.  In the
case of MACS J0025, the lack of returning analogs may be related to
its higher redshift of observation---there may not be enough cosmic
time for subclusters of the appropriate mass to form, pass through
pericenter and then apocenter, and then return to the observed
\dproj. A strength of the analog method over staged simulations (in
which two smooth cluster profiles are set up, then collided) is that
this cosmological context is naturally taken into account.

\begin{figure}
\centerline{\includegraphics[width=\columnwidth]{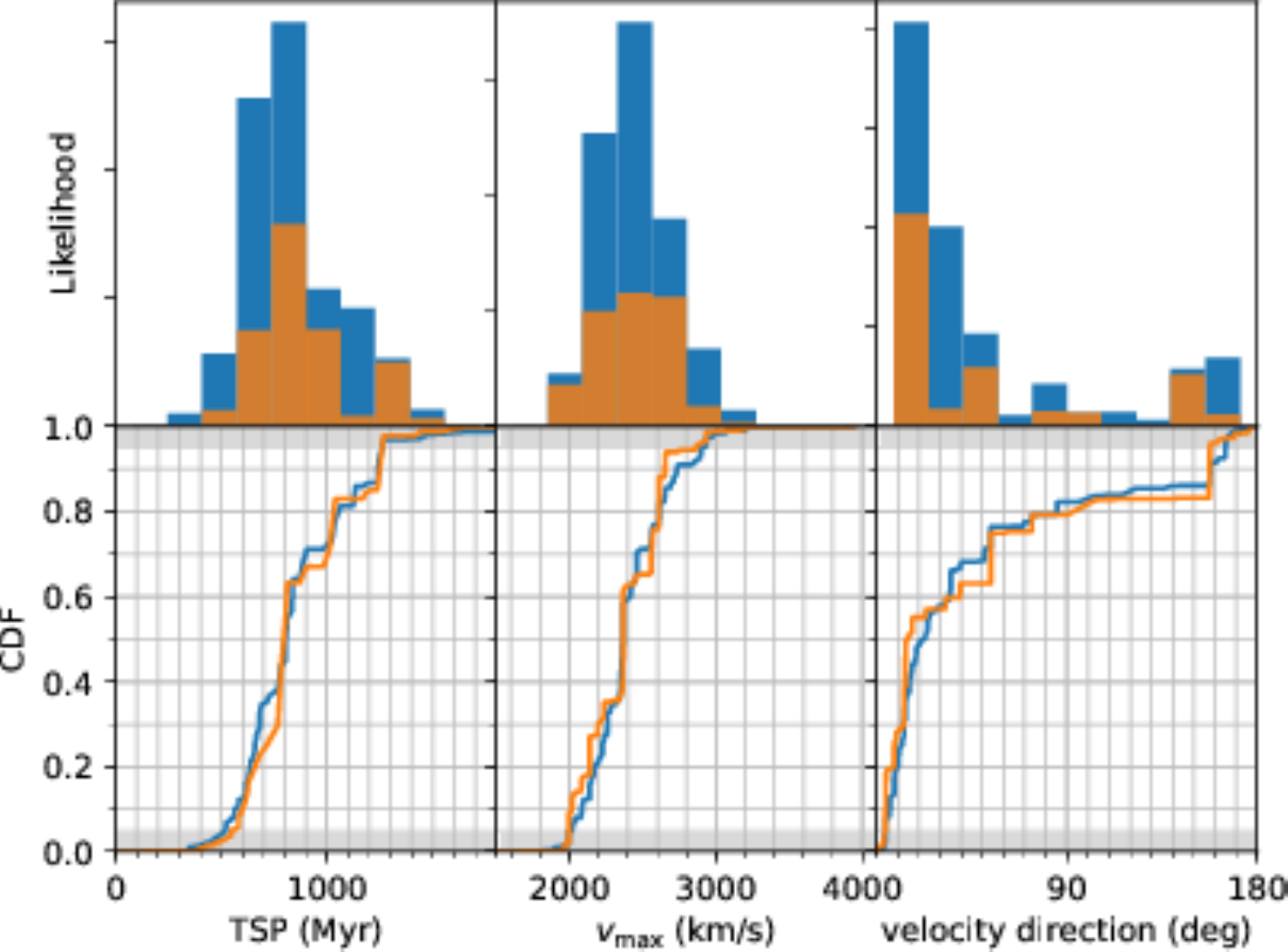}}
\caption{Likelihoods (top row) and CDFs (bottom row) for time since 
  pericenter (left column), \vmax\ (middle column), and velocity angle $\varphi$
  (right column) for ZwCl 0008.8+5215. Blue indicates all analogs, and orange 
  indicates the subset with pericenter distance $<150$ kpc. The 
  velocity angle ranges from zero for radially outbound to 180$^\circ$
  for radially inbound.}
\label{fig-0008-sextet}\end{figure}

\begin{figure}
\centerline{\includegraphics[width=\columnwidth]{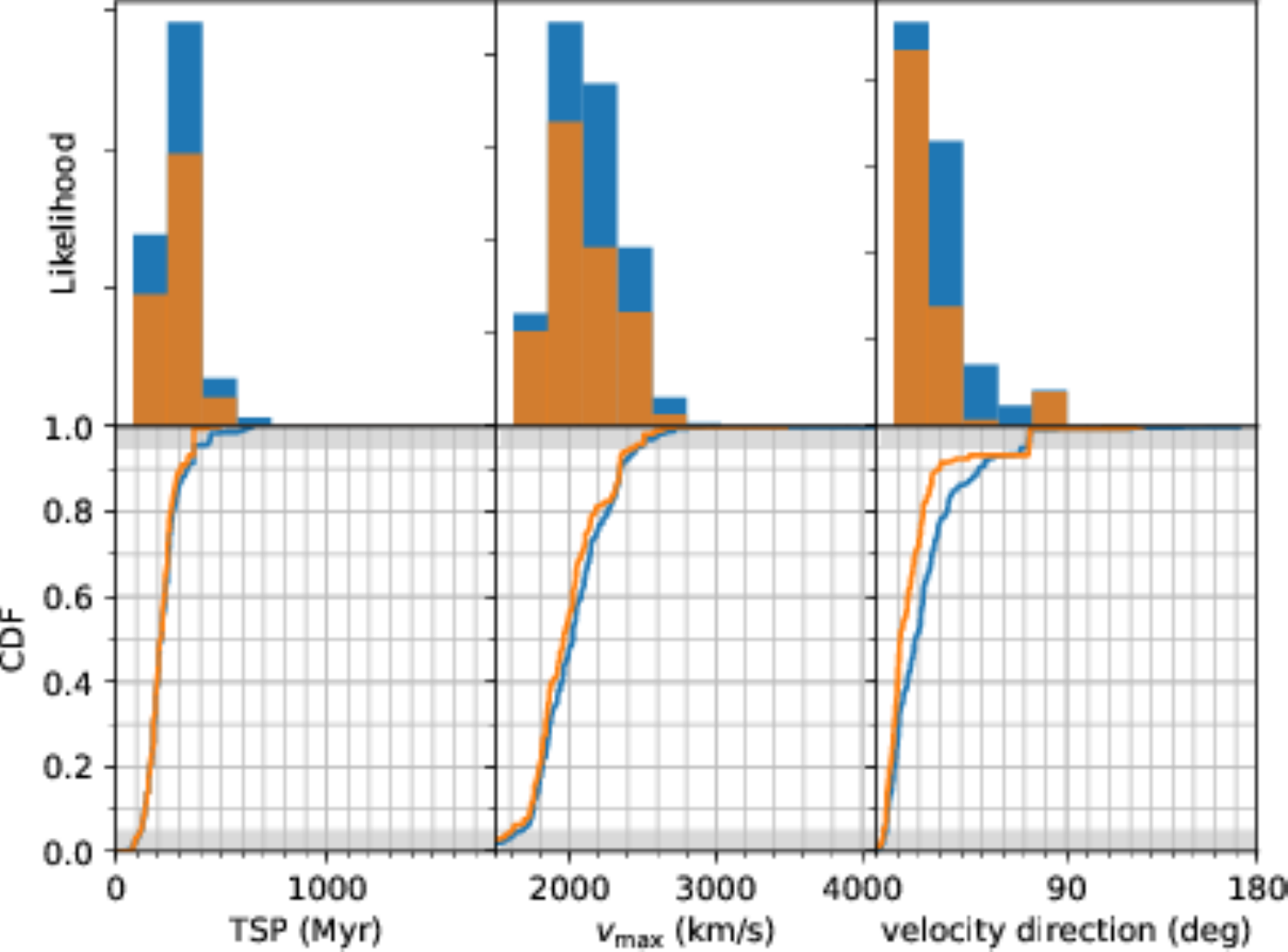}}
\caption{As for Figure~\ref{fig-0008-sextet}, but for MACS J0025.4--1222.}
\label{fig-0025-sextet}\end{figure}

We now present quantitative estimates of the three dynamical
quantities TSP, \vmax, and $\varphi$, weighted by analog likelihood.
Figure~\ref{fig-0008-sextet} focuses on \zwate, with likelihoods in
the upper panels and CDFs in the lower panels.  From left to right the
panels show TSP, \vmax, and $\varphi$.  Figure~\ref{fig-0025-sextet}
echoes this arrangement for MACS J0025.

We first verify that the results are robust to uncertainty in the
pericenter distance. Smaller pericenter distance corresponds to a more
plunging trajectory, which is more likely to strip gas and presumably
to reach a higher speed.  Although the gas morphologies in our
observed sample indicate close pericenters, we do not have
quantitative priors so we plot results for all analogs in blue, and
for the subset with pericenter distance $<150$ kpc in orange.  This is
potentially an informative cut because the orange subset includes the
most likely pericenter distance for \zwate\ based on comparison of
hydrodynamic simulations with X-ray observations
\citet{MolnarZwCl0008}: $143.5\pm6.5$ kpc (S. Molnar, private
communication).  Hence the agreement between this subset and all
analogs (up to 300 kpc, our initial selection criterion) indicates
that such specialized knowledge of the pericenter distance is
unnecessary, at least for TSP and \vmax. The $\varphi$ distribution
for MACS J0025 is somewhat sensitive to the pericenter cut, but if the
primary use of $\varphi$ is to separate models into outbound
($\varphi<90^\circ$) and returning ($\varphi>90^\circ$) then the impact
of this cut is negligible.

Because the 150 kpc cut on pericenter distance does not sway the
results but does reduce the number of analogs, we use only the initial
300 kpc selection criterion. The impact parameter is a related
quantity more commonly quoted by simulators, defined at early times
(large separations) as the component of the separation vector
perpendicular to the velocity vector. As a rule of thumb, the impact
parameter is $\gtrsim 3$ times larger than the pericenter distance
\citep{Zhang2016PericenterImpactParameter}. Hence our pericenter cut
allows impact parameters of $\gtrsim 1$ Mpc, which easily encompasses
the range suggested by hydrodynamic simulations of the systems we
consider.

% convert '../Code/ZwCl0008.pdf[6]' -flatten -trim +repage ZwCl0008-sextet.pdf

% convert '../Code/MACSJ0025.pdf[6]' -flatten -trim +repage MACSJ0025-sextet.pdf

With this final selection criterion established, we can read
confidence intervals (CI) off Figures~\ref{fig-0008-sextet} and
\ref{fig-0025-sextet}.  We find that $616<\rm{TSP}<1130$ Myr for
\zwate, and $152<\rm{TSP}<288$ Myr for MACS J0025, at 68\%
confidence. This confirms the qualitative impression from
Figures~\ref{fig-ZwCltrajectories} and \ref{fig-MACStrajectories} that
TSP must be greater for \zwate.  Regarding \vmax, we find
$2139<\vmax<2653$ km/s at 68\% confidence for \zwate\ and
$1794<\vmax<2315$ km/s for MACS J0025. In other words \zwate\ is
likely the faster merger, but the ranges do overlap.

Turning to the rightmost columns of Figures~\ref{fig-0008-sextet} and
\ref{fig-0025-sextet}, we find that both systems are likely to be
outbound ($\varphi<90^\circ$), at 82\% confidence for \zwate\ and
$99.6\%$ confidence for MACS J0025.  The strong preference for outbound
systems is striking, given that every outbound system at a given
separation eventually becomes a returning system at the same
separation.  Due to dissipation, however, the speed when returning at
the same separation must be lower. This effect may account for the
power of the analogs to discriminate between outbound and returning
phases. To test this contention we ran hypothetical versions of
\zwate\ with \dv\ successively incremented by 100 km/s: the confidence
that the system is outbound increased by about 1\% for each
increment. Hence, the observed relative speed does affect the
outbound/returning inference, but only slightly.

The redshift of observation also matters. As mentioned above, at
higher redshifts there may not be enough cosmic time for subclusters
of the appropriate mass to form, pass through pericenter and then
apocenter, and then return to the observed \dproj. We tested this by
placing the \zwate\ observations at a redshift matching MACS J0025
($z=0.59$ rather than 0.10); the confidence in the outbound model
jumps up to 95\%. For small perturbations in redshift, however, this
effect can be obscured by variations in the analog systems
selected. (This is not a source of noise when perturbing \dv\ because
the same analog systems are used, albeit with perturbed likelihoods.)

In this paper we use $\varphi$ as a binary outbound/returning
indicator, but it may also probe orbital eccentricity as
follows. Purely radial orbits will cause the likelihood peak at
$\varphi=0$ to be quite sharp, whereas the inclusion of eccentric
orbits will broaden this peak. This is closely related to the question
of pericenter distance, which as discussed above may require 
extending this method to include hydrodynamical simulations.

{\it Comparison with staged hydrodynamic simulations of \zwate.}
\citep[][hereafter MB18]{MolnarZwCl0008} recently performed a suite of simulations of
this system.  For a consistent comparison, we rerun the analog method
with the modestly higher masses (7 and $5\times10^{14}$ M$_\odot$) and
modestly lower \dproj\ they adopted.  The two methods agree that the
system is outbound, but the hydrodynamics support this conclusion at
high confidence while the analog method yields only 85\% confidence.
For a consistent comparison with their results we consider only
outbound analogs in the following:
\begin{itemize}
\item For \vmax\ we find a 68\% (90\%) CI of 2020--2561 (1829--2836)
  km/s. This is substantially lower than the MB18 value of 3515 km/s,
  which corresponds to our 99.9999\% confidence upper limit. It is
  possible that the analog speeds are underestimated by more than the
  100--200 km/s we have associated with halo particle
  misidentification.  It is also possible that substructure and
  large-scale structure, which are missing from the staged
  simulations, prevent \vmax\ from reaching the high values seen in
  staged simulations. To further explore this issue, it will be
  instructive to apply the analog method to a cosmological simulation
  with hydrodynamics.
\item For TSP we find a 68\% (90\%) CI of
313--736 (205--990) Myr, both encompassing the MB18
value of 428 Myr. Their value is slightly lower than the center of our range,
which is likely related to their higher speed.
\item Our viewing angle results are consistent. We find the separation
  vector is $\ge 73$ ($\ge 61$) degrees from the line of sight at 68\%
  (90\%) confidence, while MB18 find that 61$^\circ$ best matches the
  X-ray morphology. 
\end{itemize}

Note that confidence intervals are not given by MB18
because hydrodynamic simulations are expensive---after doing
simulations that bracket a range of parameters to find the best fit,
it is infeasible to do many more with slightly perturbed values to
support confidence intervals.  A possible further development
would be to combine the analog and hydro methods by hydrodynamically
resimulating a representative suite of analogs.

\subsection{Merging Cluster Collaboration Gold Sample}\label{sec-gold}

\citet{MCCsampleanalysis} examined a sample of 29 radio-selected
merging clusters, including analog viewing angle constraints, and
identified a gold subsample of eight cleanly bimodal systems.  These
are listed in the first block of Table~\ref{tab-clusterinputs}. For
some clusters noted, only dynamical masses are available but these are
biased high in a merger by a factor of two or more
\citep{Pinkney96,Takizawa10}. Mass overestimates trigger two
difficulties with the analog method. First, in {\it any} type of
dynamical inference, the timescales will be biased low, and the
pericenter speeds biased high, because higher masses are associated
with greater accelerations from the time of pericenter to the time of
observation. However, times and speeds depend sublinearly on the
  mass ($\sqrt{M}$ dependence for point masses) so this does not
  necessarily result in a large bias. The more difficult issue arises
  when the nominal masses are very high {\it and} have very small
  nominal uncertainties: few analogs will be found and the single
  highest-mass analog will dominate the likelihood.  This problem is
  particularly acute for MACS J1149.5+2223 and A3411, with nominal
  dynamical masses of 37.6 and 32$\times10^{14}$ M$_\odot$
  respectively, nominally $\approx 10$ times the uncertainty.

  We stress that for systems with high {\it lensing} masses (unlikely
  to be biased high in a merger) the proper way to find more analogs
  is to simulate a larger volume. Dynamical masses, however, must be
  debiased before searching for analogs. We take guidance from studies
  of the two systems mentioned above. First, \citet{Applegate14WtG}
  did a weak lensing study of all mass within 1.5 Mpc of MACS
  J1149.5+2223.  This was done before MACS J1149.5+2223 was recognized
  to be bimodal, but the 1.5 Mpc radius includes both
  subclusters. They found $14\times10^{14}$ M$_\odot$, almost exactly
  half the dynamical mass and in line with the general bias
  studies of \citet{Pinkney96} and \citet{Takizawa10}.  Second, a
  recent X-ray analysis of A3411 (Andrade-Santos {\it et. al},
  submitted) finds $M_{500,Y_X} = 7.1 \times 10^{14}$ M$_\odot$, which
  extrapolates to a total system mass
  $M_{\rm vir}\approx M_{200} \approx 10^{15}$ M$_\odot$. This is
  somewhat less than half the total dynamical mass, again in line with
  the general bias studies cited.  The dynamical mass estimates do
  help apportion the total mass into subcluster masses, because the
  bias should be similar for both subclusters. Hence, where subcluster
  lensing masses are unavailable we use half the subcluster dynamical
  mass, retaining the nominal uncertainties.

\begin{figure}
\centerline{\includegraphics[width=\columnwidth]{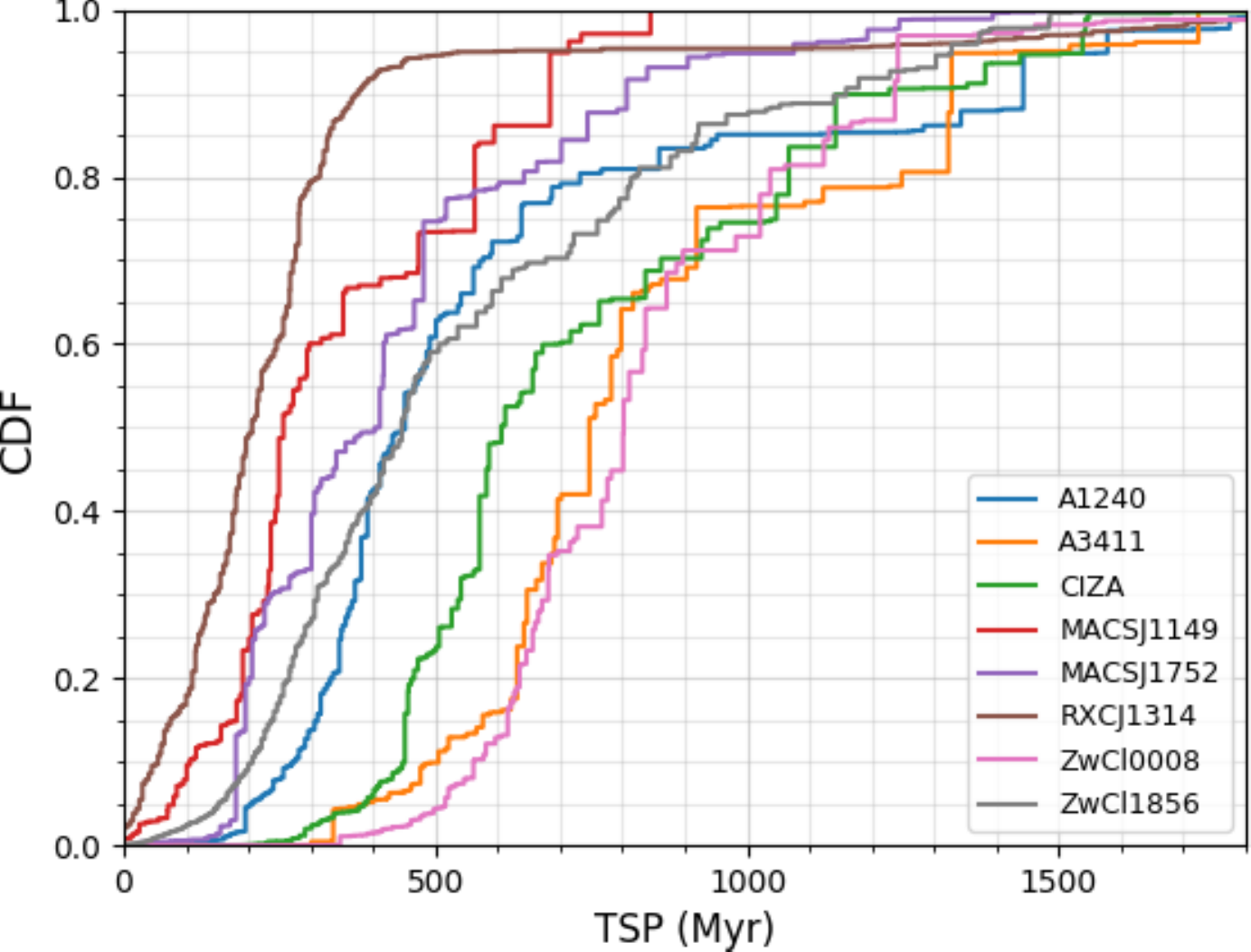}}
\caption{Cumulative distribution functions for time since pericenter
  for the gold sample defined by \citet{MCCsampleanalysis}.}
\label{fig-goldTSP}\end{figure}

Figure~\ref{fig-goldTSP} shows the TSP constraints, in the form of
CDFs, for these eight systems. This provides a clear impression of how
to rank the systems from youngest to oldest.  To quantify this we
define the median analog TSP for a system as the TSP at which that
system's CDF crosses 0.5.  RXC J1314.4-2515 is the youngest system
with a median analog TSP of 206 Myr. (For comparison, the median
analog TSP for MACS J0025, the younger system in \S\ref{sec-illpair},
is 216 Myr.)  This is followed by MACS J1149.5+2223 at 256 Myr, then a
group of three (A1240, MACS J1752.0+4440, and ZwCl 1856.8+6616) at
400--450 Myr, then CIZA J2242.8+5301 at 608 Myr. Finally, the oldest
two systems, at 750--800 Myr, are A3411 and \zwate.

The individual rankings cannot be established at high confidence
because the CDFs overlap, but there is a clear distinction between
young, middle-aged, and old. For example, the 68\% CI for RXC
J1314.4-2515 (90--327 Myr) barely overlaps that of the middle-aged
system A1240 (316--946 Myr).  These distinctions will become even
clearer in cases where prior information can rule out either the
outbound or returning phase, as follows. The returning models
constitute the long tail toward high TSP, which typically occupies the
top 5--20\% of the CDF.  This is very clear in the case of RXC
J1314.4-2515, where the last 5\% of the analog likelihood is
$\approx1.5$ Gyr older than the first 95\%; in other words, it is seen
soon after first pericenter or soon before second pericenter, but not
in between. In the returning model, the shock would have traveled very
far out and would likely be undetectable, but in fact it is detectable
and close to the subclusters \citep{Venturi2013,MazzottaRXCJ1314XMM}.
This may justify use of outbound analogs only; this has only a minor
effect on the median age but would narrow the 95\% and 99\% CI
considerably. Another example is A3411, where a few returning analogs
have a high likelihood, yielding the vertical line segments at high
TSP in Figure~\ref{fig-goldTSP}.  If these could be ruled out on the
basis of X-ray morphology (Andrade-Santos {\it et. al}, in
preparation), much stricter upper limits could be placed on TSP even
as the median analog TSP would fall only modestly, from 755 Myr to 696
Myr.  

%python3 plotallcdf.py A1240TSPCDF.txt A3411TSPCDF.txt CIZATSPCDF.txt MACSJ1149TSPCDF.txt MACSJ1752TSPCDF.txt RXCJ1314TSPCDF.txt ZwCl0008TSPCDF.txt ZwCl1856TSPCDF.txt 
% convert ~/Figure_1.png -flatten -trim +repage goldTSP.png

\begin{figure}
\centerline{\includegraphics[width=\columnwidth]{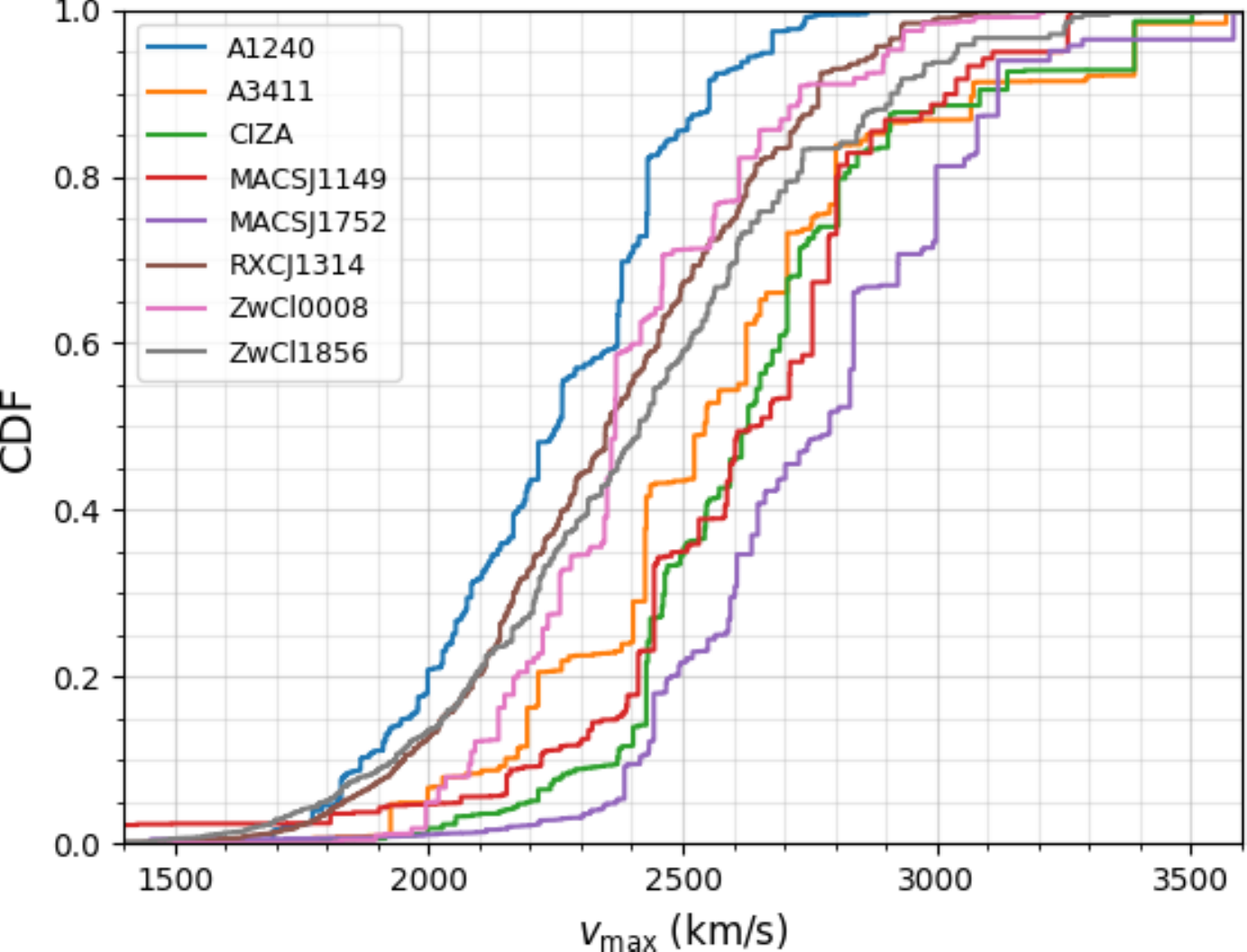}}
\caption{Cumulative distribution functions for \vmax\ for the gold
  sample defined by \citet{MCCsampleanalysis}.}
\label{fig-goldvmax}\end{figure}

Figure~\ref{fig-goldvmax} shows the \vmax\ CDFs. The two illustrative
systems considered in \S\ref{sec-illpair} would both be considered
slow in this context: MACS J0025 was found to be about 350 km/s slower
than \zwate, which is seen here in the slower half of the gold sample.
The median analog \vmax\ ranges from 2254 km/s for A1240 to 2790 km/s
for MACS J1752.0+4440. This is a small range compared to the
$\approx900$ km/s spanned by the 90\% CI of a typical system; in other
words, the differences in \vmax\ from system to system are not highly
significant.  Nevertheless, the higher-mass systems do tend to have
higher \vmax\ as one would expect from dynamics.  There is no apparent
relationship between the TSP ordering seen in Figure~\ref{fig-goldTSP}
and the \vmax\ ordering seen here.

Note that RXC J1314.4-2515 has a remarkably high observed \dv\ (1498
km/s) yet is unremarkable, even a bit low, in terms of \vmax. Part of
the explanation lies in projection effects; we agree with WCN18 and
\citet{MCCsampleanalysis} that the most likely viewing angle is around
45$^\circ$, which exposes a larger fraction of its current 3-D
velocity than do other systems.  We quantify this by tabulating the
median analog 3-D velocity at the time of observation. We find 1827
km/s, which is still roughly a factor of two higher than for most
other systems considered here. Hence projection cannot be the only
factor behind the high observed \dv.  The low TSP provides a second
factor: this system has had less time to slow down since pericenter.
This provides a lesson that the observed \dv, and even its
deprojection, should not be used as a proxy for merger speed.

% python3 plotallcdf.py A1240vmaxCDF.txt A3411vmaxCDF.txt CIZAvmaxCDF.txt MACSJ1149vmaxCDF.txt MACSJ1752vmaxCDF.txt RXCJ1314vmaxCDF.txt ZwCl0008vmaxCDF.txt ZwCl1856vmaxCDF.txt
% convert ~/Figure_1.png -flatten -trim +repage goldvmax.png

\begin{figure}
\centerline{\includegraphics[width=\columnwidth]{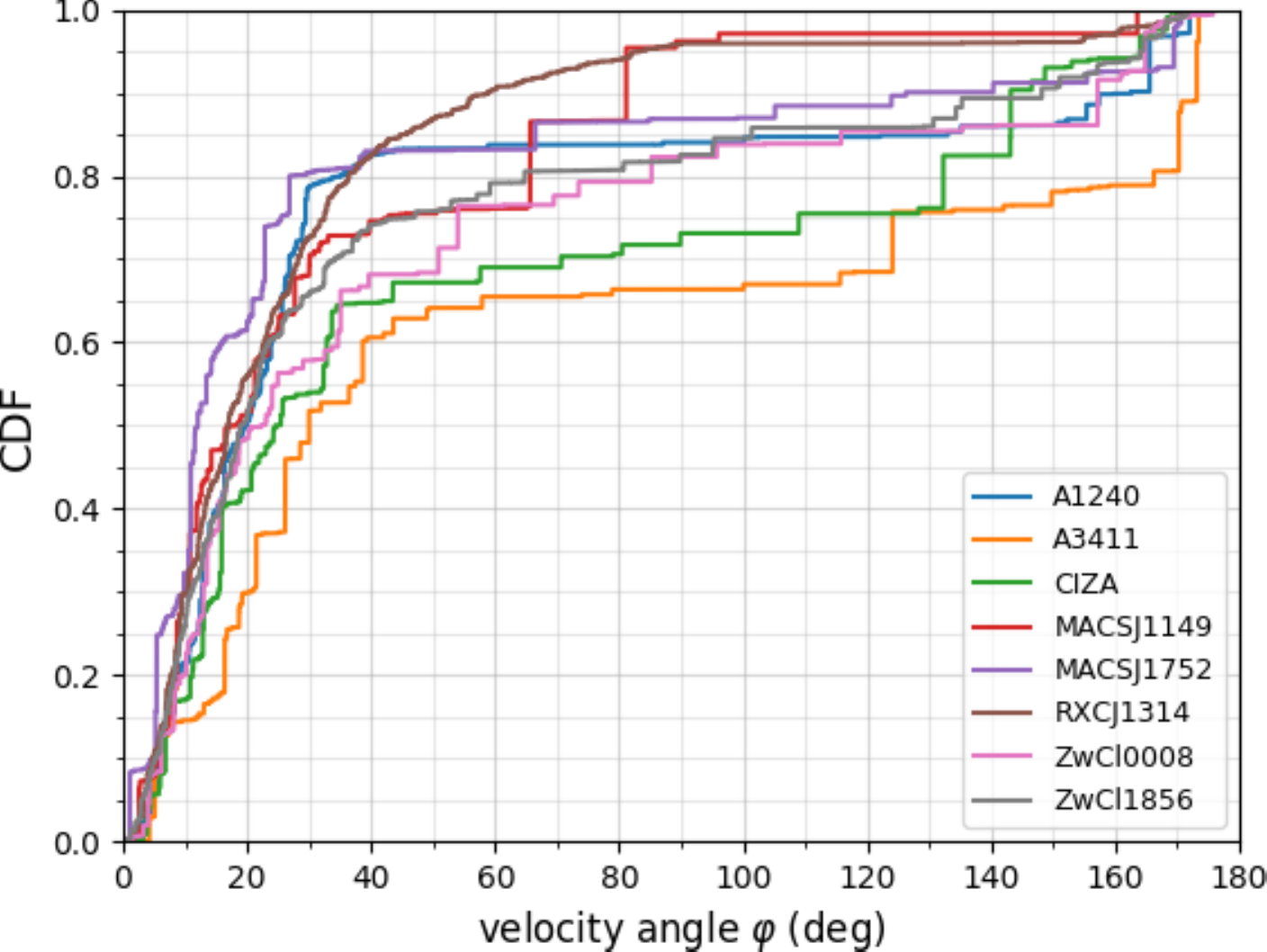}}
\caption{Cumulative distribution functions for $\varphi$ for the gold
  sample defined by \citet{MCCsampleanalysis}.}
\label{fig-goldvarphi}\end{figure}

Figure~\ref{fig-goldvarphi} shows the $\varphi$ CDFs for the same
sample. The system most likely to be in the returning phase
($\varphi>90^\circ$) is A3411, but even here the likelihood of being
outbound is great (66\%). CIZA J2242.8+5301 has a 72\% likelihood,
and all other systems have 82\% or greater likelihood, of being in the
outbound phase. Higher-redshift systems tend to be more likely
outbound compared to lower-redshift systems, which supports the
discussion of this effect in \S\ref{sec-illpair}.  As discussed above,
X-ray morphology or shock position may be more powerful ways to
determine the phase; if so, restricting analogs to the correct phase
will provide tighter constraints on other quantities, especially TSP.

% python3 plotallcdf.py A1240angCDF.txt A3411angCDF.txt CIZAangCDF.txt MACSJ1149angCDF.txt MACSJ1752angCDF.txt RXCJ1314angCDF.txt ZwCl0008angCDF.txt ZwCl1856angCDF.txt
% convert ~/Figure_1.png -flatten -trim +repage goldvarphi.png

{\it Comparison with staged hydrodynamic simulations of CIZA
  J2242.8+5301.} \citet{Molnar2017CIZA} recently performed a suite of
simulations of this system.  For a consistent comparison, we rerun the
analog method with their lower subcluster mass estimates (5.0 and
$3.9\times10^{14}$ M$_\odot$).  We agree (with 76\% confidence) that
the system is outbound.  We adopt this conclusion and consider only
outbound analogs in the following:
\begin{itemize}
\item For \vmax\ we find a 68\% (90\%) CI of 2029--2661 (1916--2746)
  km/s. \citet{Molnar2017CIZA} do not list their maximum speed, but it
  must be substantially higher than this because their model already
  has a relative speed of 2500 km/s when the two virial radii first
  touch (fully 700 km/s faster than they found for \zwate). This
  reinforces the notion that the analogs may be underestimating
    the maximum speed by more than the 100--200 km/s described in
    \S\ref{sec-method}, and/or that staged simulations may
    overestimate the maximum speed because they lack substructure and
  large-scale structure.
\item For TSP we find a 68\% (90\%) CI of 425--767 (291--1052) Myr,
  consistent with the \citet{Molnar2017CIZA} value of 0.4 Gyr. Their
  value is definitely lower than the center of our range, which is
  again likely related to their higher speed.
\item \citet{Molnar2017CIZA} suggest that at the time of observation
  the separation vector is 75$^\circ$ from the line of sight. We agree,
  with 68\% (90\%) confidence lower limits of 70 (59) degrees. 
\end{itemize}

\section{Comparison to MCMAC}\label{sec-mcmac}

\begin{figure}
\centerline{\includegraphics[width=\columnwidth]{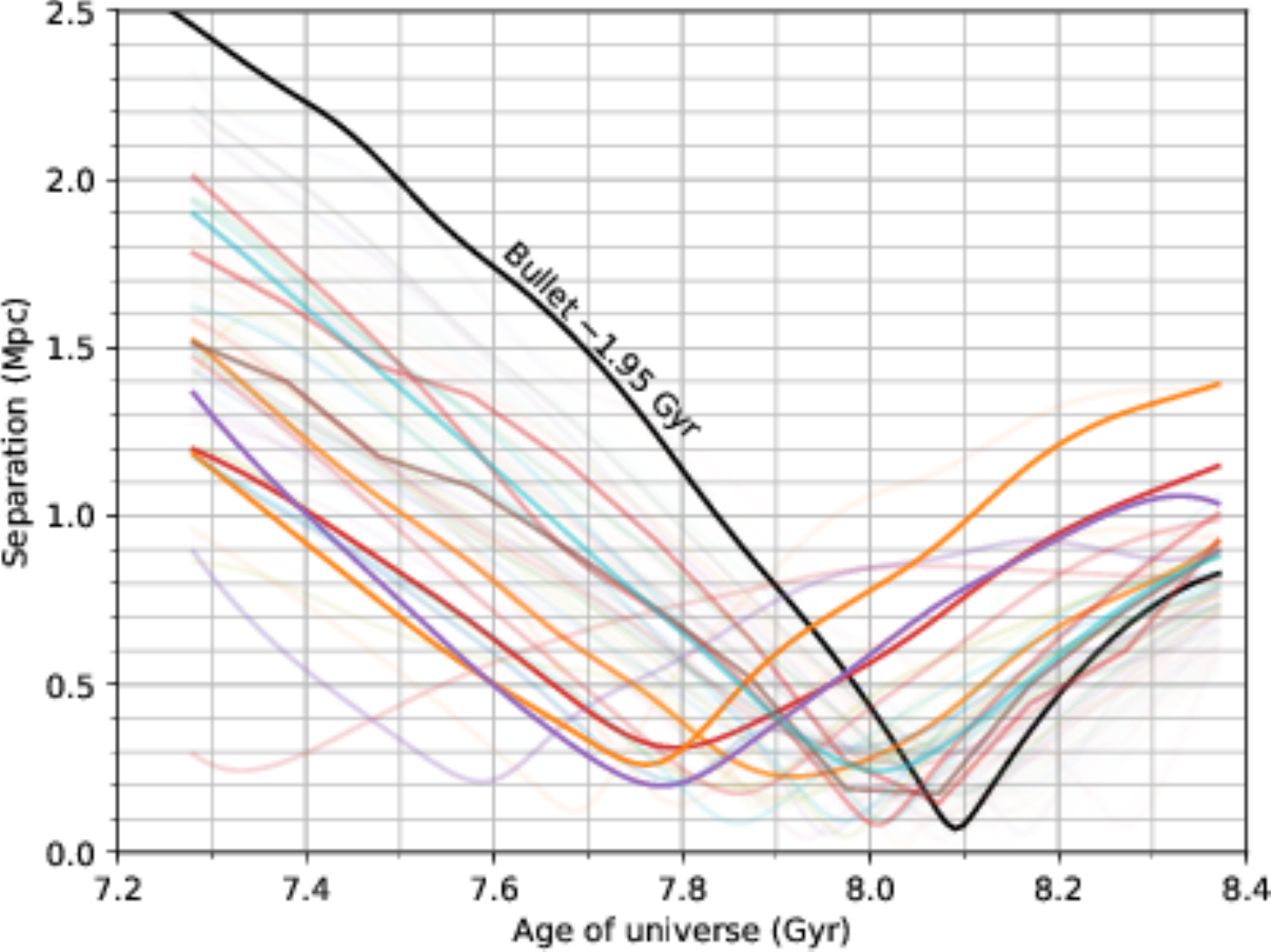}}
\caption{Colored curves show physical separation versus time for
  analogs of the Musketball cluster, with opacity indicating the
  likelihood. The black curve is the dominant analog for the Bullet
  cluster, shifted in time so the observed states (the ends of the
  curves) can be compared.}
\label{fig-bullettraj}\end{figure}

\begin{figure*}
\centerline{\includegraphics[width=\textwidth]{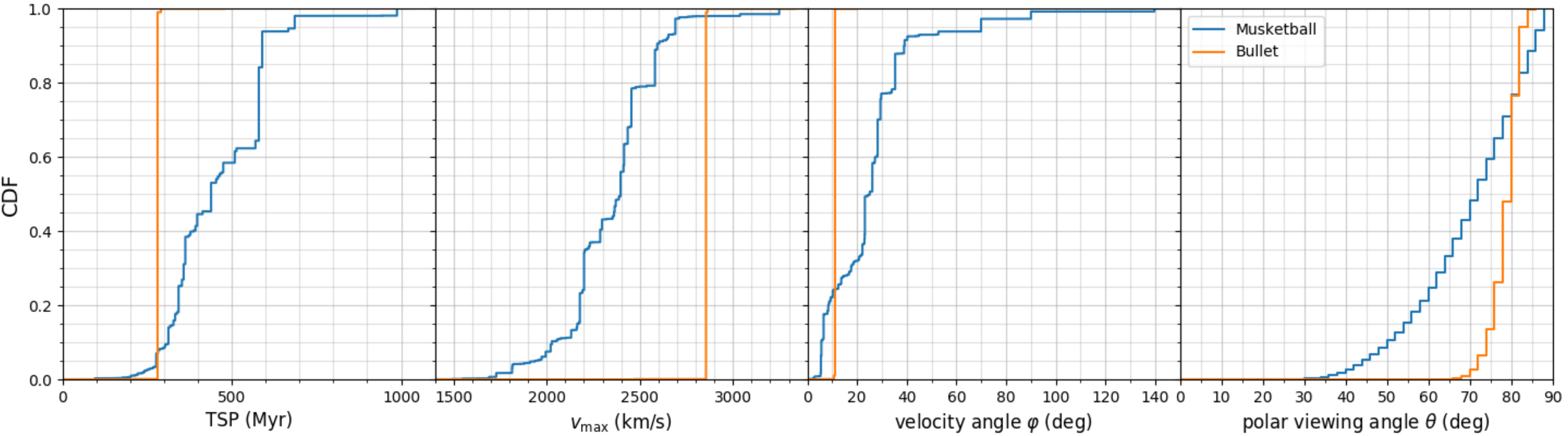}}
\caption{Analog constraints on the properties of the Musketball are
  shown in blue. Properties of the single dominant Bullet analog are
  shown in orange.}
\label{fig-musketvbullet}\end{figure*}

We now compare our results to those of \citet{Dawson2012}, hereafter
D13, who analyzed the Bullet cluster (1E 0657-558) and the older and
slower Musketball cluster (DLSCL J0916.2+2951) with MCMAC. We adopt
the \dv\ and mass values used by D13, which came from observations by
\citet{Barrena2002Bulletredshifts} and \citet{BradacBulletLensing2006}
for the Bullet, and by \citet{Dawson11} for the Musketball. D13 drew
\dproj\ values from these sources as well; we convert those to angular
values using his assumed cosmology, and then back to kpc using the
BigMDPL cosmology, which increases the physical values by about
3\%. We display trajectories of the analogs in
Figure~\ref{fig-bullettraj} and dynamical inferences in
Figure~\ref{fig-musketvbullet}.

We begin with the Musketball. We find a 68\% CI of 333--590 Myr for
TSP, compared to 700--2400 Myr found by D13. We attribute this to the
effect explained in \S\ref{sec-intro}: because the MCMAC model
examines only radial trajectories, it is forced to explain nonzero
\dv\ by reducing $\theta$ from its {\it a priori} most likely value of
$90^\circ$. In contrast, in BigMDPL analogs the relative velocity
vector usually has a component perpendicular to the separation vector,
which readily allows $\theta=90^\circ$ models (i.e. the separation
vector is in the plane of the sky) unless the observed \dv\ is more
than a few hundred km/s. The Musketball \dv\ of $630\pm330$ km/s is
enough to broaden the likelihood peak away from $\theta=90^\circ$,
yielding a plateau across the range 78--90$^\circ$. This contrasts
with the D13 estimate that $\theta\approx48^\circ$ is most likely.
(In terms of 68\% CI, we find $\theta\ge 65^\circ$, whereas D13 found
23--62$^\circ$.)  Hence MCMAC infers a substantially larger 3-D
separation at the time of observation, which in turn requires more TSP
to reach that separation.

The two methods agree on \vmax: we find a 68\% CI of 2178--2582 km/s,
while D13 found 2000--2500 km/s. (As noted in \S\ref{sec-method}, the
maximum velocity recorded in the BigMDPL snapshots, which we quote
here, may underestimate the maximum simulated velocity by 100--200
km/s.)

Analogs provide two additional quantities that cannot be provided by
MCMAC. First, the analytical model in MCMAC can make no distinction
between outbound and returning phases. The D13 TSP we quoted assumes
the outbound phase, but D13 provided a second TSP that assumes the
returning phase: 2.0--7.2 Gyr. Our analogs favor the outbound phase at
97\% confidence. (The few returning analogs we do find have
TSP$\approx 0.9$ Gyr, indicating that our analogs have shorter periods
than in the D13 calculation.) Second, MCMAC assumes zero pericenter
distance, but we are able to extract the pericenter distance of the
analogs. We find a median analog pericenter distance of 180 kpc, with
negligible probability that the pericenter distance is less than 20
kpc.  This is a case where improved constraints on pericenter
distance, perhaps from hydrodynamic simulations, could improve the
analog constraints. The Musketball does have analogs with large
pericenter distances (180--280 kpc) and these favor TSP$\approx$600
Myr. If these large pericenter distances could be ruled out based on
the dissociative X-ray morphology, the TSP estimates would tighten
around 350 Myr, near the lower end of our current range. There would
be little effect on the \vmax\ and outbound/returning estimates in
this scenario.

For the Bullet, D13 offers two sets of results: a default set, and one
with an additional prior limiting the TSP, which is justified based on
the observation of transient X-ray effects such as the shock front and
increased X-ray luminosity (compared to expectations from the lensing
mass).  This has the effect of reducing the 68\% CI for TSP from
0.3--1.1 Gyr to 0.3--0.5 Gyr, assuming the outbound phase. We will
compare our results to the latter set of D13 results.

We find that one analog dominates, providing 98.9\% of the weight,
because the observational uncertainties listed by D13 are quite
small. Rather than attempt to give confidence intervals, we focus on
this one dominant analog.  This analog is consistent with the D13
estimates. Its TSP is 281 Myr, a bit below the D13 68\% CI but well
within his 95\% CI of 0.2--0.6 Gyr. The analog has a \vmax\ of 2863
km/s, well within the D13 68\% CI of 2600--3300, and a most likely
$\theta$ of 81$^\circ$, which is slightly outside the D13 68\% CI of
52--74$^\circ$, but understandable on the basis of D13's radial
assumption. The analog also has a pericenter distance of 74 kpc and a
current angle of 11$^\circ$ between the velocity and separation
vectors.

The next most likely analog carries almost 1\% of the weight and is
quite similar: TSP\ $=291$ Myr, $\vmax=2871$ km/s, and
$\varphi=10^\circ$, but with a larger pericenter distance of 136 kpc.
Providing confidence intervals for the Bullet would require finding
additional analogs that can compete with the dominant analog. If the
fairly strict uncertainties on observed quantities are correct in this
well-observed system, this would in turn require simulating a larger
volume.  MCMAC, in contrast, is always able to provide smooth formal
confidence intervals because analytical models can always be
perturbed, while the analog method is limited by the discrete
number of available analogs.  This weakness of the analog method could
potentially be ameliorated by using ``genetically modified''
simulations \citep{2018MNRAS.474...45R} to produce larger quantities
of relevant analogs.

Even a single analog can be viewed from a range of angles, making
confidence intervals on $\theta$ mathematically possible. The dominant
Bullet analog, for example, yields a formal 90\% CI of
73--84$^\circ$. Such confidence intervals do not reflect
marginalization over a cosmologically motivated range of impact
parameters, so we recommend caution in this regard. This dominant
analog nevertheless agrees with D13 and WCN18 in putting the
separation vector 10--20$^\circ$ from the plane of the sky.

Finally, we note that hydrodynamic simulations of the Bullet
  cluster \citep{Springel2007} prefer a younger model with TSP of 180
  Myr.  The observed separation is achieved quickly with
  $v_{3D}(t_{\rm obs})\approx 2700$ km/s, versus our dominant analog
  with a {\it maximum} speed of 2863 km/s and a current speed of 1965
  km/s.  This continues the pattern in which staged hydrodynamic
  simulations produce higher speeds than our analogs. Further work,
  perhaps with cosmological hydrodynamical simulations, will be needed
  to clarify the source of this discrepancy.

\section{Summary and Discussion}\label{sec-discussion}

We identified analogs of observed merging galaxy clusters in the
BigMDPL cosmological n-body simulation based only on subcluster
masses, relative line-of-sight speed, and projected separation.  We
then extracted dynamical properties of the analogs such as time since
pericenter, maximum relative speed, and merger phase (outbound or
returning) at the time of observation.  Table~\ref{tab-results} lists
highest probability density confidence intervals for TSP
and \vmax, as well as the percentage confidence that the system is
outbound.

\begin{table*}
\centering 
\begin{tabular}{lccccc}
Cluster & TSP (Myr, 68\% CI) & TSP (95\% CI) & \vmax (km/s, 68\%) & \vmax
              (95\%) & \% outbound\\ 
\hline 
\multicolumn{6}{c}{Merging Cluster Collaboration Gold Sample}\\
\hline 
Abell 1240 &195--577 &195--1583 &1979--2466 &1758--2683 &84\\
Abell 3411 &  476--971 &258--1438 &2194--2808 &1926--3502 &66\\
CIZA J2242.8+5301 &  378--937 &378--1576 &2403--2808 &2135--3502 &72\\
MACS J1149.5+2223&   181--569 &15--734 &2391--2888 &1903--3264 &96\\
MACS J1752.0+4440 &   180--491 &135--1113 &2444--3034 &1838--3264 &87\\
RXC J13144.4--2515&   100--328 &0--627 &2018--2653 &1786--2934 &96\\
ZwCl 0008.8+5215  & 516--897 &411--1267 &2020--2461 &1995--2938 &81\\
ZwCl 1856.8+6616 &   50--627 &95--1373 &2036--2749 &1555--3046 &82\\
\hline 
\multicolumn{6}{c}{Other Clusters}\\
\hline 
MACS J0025.4--1222 & 132--258 &71--370 &1749--2204 &1503--2509 &100\\
DLSCL J0916.2+2951 & 333--590 &267--782 &2054--2457 &1814--2699 &97\\
1E 0657--558 & \multicolumn{2}{c}{281 (single analog)}&
                                                    \multicolumn{2}{c}{2863
                                                    (single analog)} & 100 \\
\end{tabular}
\caption{Inferred dynamical properties of merging clusters.}
\label{tab-results}
\end{table*}

Our major results are:
\begin{itemize}
\item Although the uncertainties on TSP can be in the hundreds of Myr,
  the analogs can distinguish between ``young'' mergers seen 200--280
  Myr after pericenter (MACS J1149.5+2223, RXC J1314.4-2515, MACS
  J0025 4--1222, and the Bullet), intermediate systems at 400--450 Myr
  (MACS J1752.0+4440, Abell 1240, ZwCl 1856.8+6616, and the
  Musketball) and ``old'' mergers seen 600 Myr (CIZA J2242.8+5301) or
  even 700--800 Myr (Abell 3411 and ZwCl 0008.8+5215) after pericenter.

\item All these systems are more likely to be outbound than returning,
  but in most cases the returning phase cannot be ruled out at high
  confidence based on these analog matching criteria alone. Because of
  this, 95\% confidence upper limits on TSP can be up to 1 Gyr older
  than the central values given above. If, for any given system, one
  or the other phase can be ruled out on the basis of other
  information, the limits on TSP for that system will shrink
  dramatically.  This in turn will improve inferences about
  time-dependent physical processes such as dark matter displacements
  \citep{Kim17}.

\item The probability of being in the returning phase increases with
  the cosmic time of observation. It also increases (albeit weakly) if one lowers the
  observed line-of-sight relative velocity while keeping the other
  parameters fixed. 

\item The maximum speed \vmax\ ranges from about 2000 km/s for MACS
  J0025.4-1222 to about 2800 km/s for MACS J1752.0+4440 and the
  Bullet.  The \vmax\ estimate depends mostly on the mass used for the
  analog selection, and not so much on the observed line-of-sight
  relative velocity. This is an indication that the underlying
  dynamics are being recovered, independent of viewing angle and time
  of observation. However, the maximum speed we tabulate in BigMDPL
  snapshots is underestimated by 100-200 km/s because of difficulties
  in assigning particles to halos at the time of pass-through.
\end{itemize}

Regarding the development and applicability of the method, we find:
\begin{itemize}
\item The analog method is fast, requiring only seconds of CPU time on
  a laptop computer per observed system after initial setup. It also
  naturally incorporates the effects of dynamical friction,
  substructure, and large-scale structure; and marginalizes over a
  cosmologically motivated range of impact parameters and
  trajectories.

\item The simulation we use here to demonstrate the method, BigMDPL,
  has dark matter only. This inclusion of hydrodynamics could affect
  the results stated above, but this method can easily be applied to
  bigger and better simulations as they become available.

\item For a few observed systems, the method is hampered by lack of
  analogs. Larger simulation volumes will be welcome, but we also note
  that lack of analogs can also result from unrealistically small
  nominal uncertainties on observed quantities. Hence it is important
  for observers to capture all sources of uncertainty and, ideally,
  make their full likelihood (or posterior) distributions available.

\item Analog selection could likely be improved if analogs were
  selected from a hydrodynamic simulation based on likelihood of
  matching the observed X-ray morphogy. This would particularly help
  with constraining the pericenter distance and outbound versus
  returning phase.  In a simplified version of this, we eliminated
  returning analogs of ZwCl 0008.8+5215 and CIZA J2242.5+2223 based on
  previously published comparison of staged simulations with the X-ray
  morphology \citep{MolnarZwCl0008,Molnar2017CIZA}. This in turn
  improved the upper limits on TSP (because returning analogs have
  such large TSP), which highlights the complementarity of the two
  methods.  Prior knowledge of the pericenter distance is less
  important for the clusters examined in this paper, but may prove
  useful for clusters that are 
  far from head-on \citep[e.g. Abell 115;][]{Kim2019A115} 

\item In principle, the shock location---using X-ray or radio
  observations---could also be included in the analog selection
  criteria. However, this requires higher resolution simulations than
  required for gross X-ray morphology, and may be infeasible with
  cosmological box sizes.  One workaround could be to use the basic
  analog selection technique to identify targets for zoom simulations
  with hydrodynamics, which then help refine the analog likelihoods.

  % \footnote{There are conflicting claims regarding El Gordo
  % \citep[ACT-CL
  % J0102-4915][]{Ng2015,Molnar2015Gordo,Zhang2018ElGordoHydro}. Given
  % the high redshift of this system, it is unlikely that returning
  % analogs could be found. However, the lack of BigMDPL snapshots at
  % high redshift prevents us from quantifying this.}

\item  The analogs suggest that all the systems studied here are
    outbound rather than returning, but no conclusion can be made at
    high confidence. Thus, the best way to apply the analog method
  may be to identify the phase based on X-ray morphology or shock
  position, and use this to inform the analog selection. None of
    the systems studied here are returning based on these criteria, so
    studying analogs of known returning systems could be a direction
    for future work.
\end{itemize}

To compare with the analytical MCMAC \citep{Dawson2012} method,
we analyzed the same two clusters considered in that paper, using the
same inputs.  We agree with D13 that the Musketball ((DLSCL 
  J0916.2+2951) is older and slower than the Bullet (1E 0657-558). In
  more detail:
\begin{itemize}
\item We find $\vmax\approx 2350$ km/s for the Musketball and 
$\vmax\approx 2850$ km/s for the Bullet, in agreement with D13;
however we remind readers that the true numbers must be somewhat
higher due to limitations of the BigMDPL halo catalog, and that our
Bullet results are based on a single highly dominant analog.

\item Our TSP is lower than that of D13, by a factor of a few for the
  Musketball. This is because our analogs have separation vectors
  substantially closer to the plane of the sky, hence a smaller
  current 3-D separation vector which can be reached in less time from
  pericenter.  This in turn stems from the bias identified by WCN18:
  the assumption of a radial trajectory artificially prohibits
  plane-of-sky configurations when (as for most observed systems) \dv\
  is nonzero.  We also find a lower TSP for the Bullet, but here the
  disagreement is smaller because the two methods agree on the viewing
  angle.

\item In principle the analog method can determine whether a system is
  outbound or returning, whereas in MCMAC this is a discrete
  degeneracy. The dominant Bullet analog is outbound and Musketball
  analogs favor outbound at 97\% confidence.
\end{itemize}

We also compared our results with staged hydrodynamical simulations of
\zwate\ \citep{MolnarZwCl0008} and CIZA J2242.8+5301
\citep{Molnar2017CIZA}.  We agree on the merger phase (both systems
are outbound rather than returning) and TSP, but disagree on
pericenter speed: the staged simulations are $\gtrsim 1000$ km/s
faster.  This discrepancy may be due to some combination of missing
hydrodynamic effects in BigMDPL and missing cosmological effects
(substructure, large-scale structure) in the staged simulations.  To
include all relevant effects, likely analogs should be resimulated
with hydrodynamics included.  Another possibility is that the
  analog method relies on observed \dv\ values that are
  underestimated; mistakes in assigning galaxies with spectroscopic
  redshifts to subclusters will always reduce the apparent relative
  velocity of the subclusters. To investigate this possibility,
  simulations could be used to generate mock observations of subhalo
  positions and redshifts, to be analyzed with the same subclustering
  techniques used by observers.

\acknowledgments

I thank the anonymous referee and Matthew Self for useful comments,
Sandor Molnar for discussions regarding his simulation results, and
Will Dawson for the original idea several years ago.  This work was
supported in part by NSF grant 1518246. The CosmoSim database used in
this paper is a service by the Leibniz-Institute for Astrophysics
Potsdam (AIP).  The MultiDark database was developed in cooperation
with the Spanish MultiDark Consolider Project CSD2009-00064. This work
also made use of {\tt astropy.cosmology} and Ned Wright's online
cosmology calculator \citep{Wright2006CosmologyCalculator}.

\bibliography{ms}

\end{document}